\documentclass[aps,showpacs,twocolumn,prb]{revtex4-1}
\usepackage[T1]{fontenc}
\usepackage{textcomp}
\usepackage{lmodern}
\usepackage[latin1]{inputenc}
\usepackage[swedish,english]{babel}
\usepackage{amsmath}
\usepackage{amsfonts}
\usepackage{graphicx} 
\usepackage{multirow}
\usepackage{calc,xfrac}
\usepackage{multirow}
\usepackage{color}
\usepackage{hyperref}
\usepackage{subfig}

\hypersetup{
        colorlinks = true,
        pdftitle = {}, 
        pdfauthor = {},
        pdfkeywords = {},
        linkcolor = blue,
        citecolor = blue,
        filecolor = black,
        urlcolor = magenta
}

\graphicspath{{./graphics/}}

\begin{document}

\title{Magnetic properties of (Fe$_{1-x}$Co$_{x}$)$_2$B alloys and the effect of doping by 5$d$ elements}

\author{A. Edstr\"{o}m}
\author{M. Werwi\'{n}ski}
\author{Diana Iu\c{s}an}
\author{J.~Rusz}\email[Corresponding author: ]{jan.rusz@physics.uu.se}
\author{O. Eriksson}
\affiliation{Division of Materials Theory, Department of Physics and Astronomy, Uppsala University,
Box 516, SE-751 20, Uppsala, Sweden}
\author{K.P.~Skokov}
\author{I.A.~Radulov}
\author{S.~Ener}
\author{M.D.~Kuz'min}
\author{J.~Hong}
\author{M.~Fries}
\author{D.Yu.~Karpenkov}
\author{O. Gutfleisch}
\affiliation{Functional Materials, Department of Materials Science, Technische Universit\"at Darmstadt, Darmstadt D-64287, Germany}
\author{P. Toson}
\author{J. Fidler}
\affiliation{Advanced Magnetics Group, Institute of Solid State Physics, Vienna University of Technology, Wiedner Hauptstra\ss{}e 8-10, 1040 Vienna, Austria}

\newcommand{\fecob}{(Fe$_{1-x}$Co$_x$)$_2$B}

\begin{abstract}
We have explored, computationally and experimentally, the magnetic properties of \fecob{} alloys. Calculations provide a good agreement with experiment in terms of the saturation magnetization and the magnetocrystalline anisotropy energy with some difficulty in describing Co$_2$B, for which it is found that both full potential effects and electron correlations treated within dynamical mean field theory are of importance for a correct description. The material exhibits a uniaxial magnetic anisotropy for a range of cobalt concentrations between $x=0.1$ and $x=0.5$. A simple model for the temperature dependence of magnetic anisotropy suggests that the complicated non-monotonous temperature behaviour is mainly due to variations in the band structure as the exchange splitting is reduced by temperature. Using density functional theory based calculations we have explored the effect of substitutional doping the transition metal sublattice by the whole range of 5$d$ transition metals and found that doping by Re or W elements should significantly enhance the magnetocrystalline anisotropy energy. Experimentally, W doping did not succeed in enhancing the magnetic anisotropy due to formation of other phases. On the other hand, doping by Ir and Re was successful and resulted in magnetic anisotropies that are in agreement with theoretical predictions. In particular, doping by 2.5~at.\% of Re on the Fe/Co site shows a magnetocrystalline anisotropy energy which is increased by 50\% compared to its parent (Fe$_{0.7}$Co$_{0.3}$)$_2$B compound, making this system interesting, for example, in the context of permanent magnet replacement materials or in other areas where a large magnetic anisotropy is of importance.
\end{abstract}


\maketitle

\section{Introduction}

Permanent magnets are used in a wide range of applications, for example in data storage, energy conversion, magnetic refrigeration, wind power generators, to name just a few\cite{Gutfleisch2011}. The determining properties of the quality of permanent magnet materials are the operating temperature range and, primarily, the energy product. In terms of operating temperature range, it depends on particular applications, but mostly we demand that the material remains magnetic well above room temperature. For the energy product, we generally require it to be as large as possible. Energy product is a relatively complex quantity; it depends on the structure of the material (size of crystalline grains, their shape and orientation), but from the microscopic point of view it depends mainly on the saturation magnetization $M_s$ and magnetocrystalline anisotropy energy (MAE), which is a measure of how difficult it is to rotate the magnetization direction by external magnetic field\cite{kirchmayr,sugimoto,mccallum}.

The best permanent magnet materials known today are typically utilizing rare-earth elements, most notably, neodymium in Nd$_2$Fe$_{14}$B and samarium in SmCo$_5$ magnets. Both exhibit high Curie temperatures and large energy products. Recently, however, the availability (and thus price) of the rare-earth elements became rather volatile, calling for development of replacement materials, which would use less or none of the rare-earth elements. Intense research efforts have started worldwide, revisiting previously known materials, such as Fe$_2$P\cite{Fujii,ernafe2p,leitao}, FeNi\cite{FeNi} or Fe$_{16}$N$_{2}$ \cite{fe16n2}, doing computational data mining among the large family of Heusler alloys \cite{heuslers}, exploring the effects of strain \cite{Burkert2004,Andersson2006,Warnicke2007,Yildiz2009,Costa,Neise2011,Turek2012} and doping by interstitial elements \cite{fecoc,fecocexp}, multilayers such as Fe/W-Re\cite{Bhandary}, or as a limiting case of multilayers, the L$1_0$ family of compounds \cite{l10}, or promising Mn-based systems \cite{mnbi1, mnbi2, mnbi3, mnbi4, Jian2015, Ener2015}, among others.

In this work we explore the magnetic properties of the \fecob{} alloys. In 1970 the magnetic properties of these compounds have been experimentally studied by Iga\cite{iga}, later by Coene et al.\cite{Coene} and very recently by Kuz'min et al.\cite{Kuzmin2014}. It was found that within a certain range of Co contents the material exhibits uniaxial magnetocrystalline anisotropy, with a relatively large MAE, significantly higher than pure iron or even cobalt elemental metals. To the best of our knowledge, theoretical understanding of the observations is missing.

Moreover, motivated by the works of Andersson \cite{Andersson2007} and Bhandary \cite{Bhandary}, we have attempted to further improve the MAE of the \fecob{} materials by doping them with 5$d$ transition metal elements. The mentioned works have shown that the hybridization of the exchange-split 3$d$ bands of Fe or Co with the 5$d$ bands of heavier transition metal elements may lead to a significant enhancement of the MAE, even if the 5$d$ element itself is non-magnetic or very weakly magnetic. We have thus considered low percentage substitutions of Fe or Co by the whole series of 5$d$ transition metal elements and computationally explored the effect of the substitutions on the MAE.

Together with the first principles electronic structure calculations we have performed experimental studies of \fecob{} alloys for selected Co concentrations and at the optimal concentration in terms of maximal MAE, we have synthesized single crystals with 2.5~at.\% of selected 5$d$ elements alloyed on the 3$d$ site. For these single crystals we have then measured the magnetic properties and compared them to previous results\cite{iga} and our calculations.

The structure of the manuscript is the following: In Sec.~\ref{sec:exp} we present our experimental results of synthesis and magnetic characterization of \fecob{} alloys. In Sec.~\ref{sec:dft} we present our results of computational studies of \fecob{} alloys using two different approaches to the problem of substitutional disorder. Different computational models are compared and the importance of correlation effects are discussied. Furthermore, we relate the MAE to the electronic band structure around the Fermi energy (E$_\text{F}$) since it is well known that the MAE in $d$-electron systems is determined by the electronic structure close to E$_\text{F}$\cite{Kondorskii} and in particular by the spin-orbit coupling (SOC) between occupied and unoccupied states with small difference in energy. Sec.~\ref{sec:imp} then presents theoretical and experimental results of the \fecob{} alloys doped by 5$d$ transition metal elements.

\section{Measurement of Magnetic Properties}\label{sec:exp}

\subsection{Experimental Methods}\label{sec:expmeth}

Unambiguous determination of anisotropy constants requires single crystals. Therefore, several single crystals of (Fe$_{1-x}$Co$_{x}$)$_2$B were grown as part of this work. The first stage consisted in melting a mixture of Fe and Co (both 99.99\% pure) with crystalline B of 99.999\% purity. The melting was performed in alumina crucibles placed in an induction furnace under a gauge pressure of 2 bars of pure argon. When we proceeded from the stoichiometric composition, the powder XRD patterns of \fecob{} alloys corresponded well to the CuAl$_2$-type structure but the ingots also contained some precipitants (3-4 wt.\%) rich in the $3d$ elements. To avoid the precipitants, the starting mixture was taken with a 1.5~at.\% excess of B; the so obtained ingot appeared as single-phase in powder x-ray diffraction patterns and scanning electron microscope images. 

The ingots were subsequently remelted. For better homogeneity, the melts were held for 5 minutes at 1425$^{\circ}$C in an argon atmosphere. Then the temperature was reduced slowly, at a rate of 0.1$^{\circ}$C/min, down to 1100$^{\circ}$C. At that point the power was switched off and the furnace quickly (initially at approx. 1$^{\circ}$C/min) cooled down to room temperature. The remelted ingots contained large (approx. 2-3 mm) crystalline grains. Individual grains were isolated and polished to give them the shape of a rectangular prism with edges parallel to the principal crystallographic axes. The crystals studied in more detail had the dimensions approx. 0.5$\times$3$\times$3 mm$^3$ with one of the longest edges being along [001] and the other one along [100]. X-ray back-scattering Laue patterns were used for the orientation and quality control of the crystals.

The XRD measurements were done at room temperature using a STOE Stadi P diffractometer with Mo K$\alpha_1$ radiation.  

Magnetization isotherms were measured in static magnetic fields up to 3~T at temperatures ranging between 10 and 1000~K using a Physical Property Measurement System (PPMS-14 of Quantum Design). The magnetic field was applied either in the magnetization direction [001] or along [100]. Test measurements were also performed for the direction [110] and the resulting curves proved indistinguishable from the corresponding [100] curves. 

\subsection{Magnetization and Magnetocrystalline Anisotropy}

\begin{figure}[hbt]
	\centering
	\includegraphics[width=8.5cm]{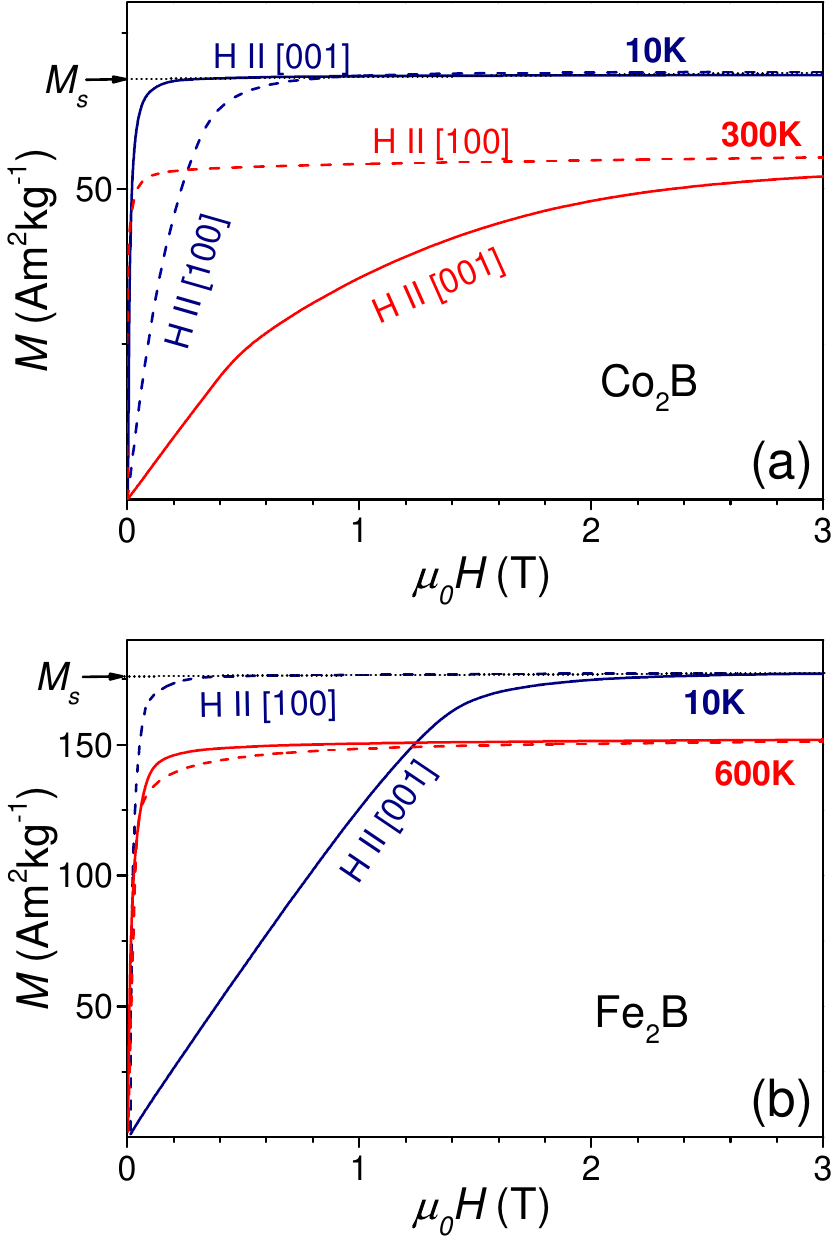}
	\caption{Magnetization curves of (a) Co$_2$B and (b) Fe$_2$B single crystals measured along [001] (dashed lines) and [100] (solid lines).}
	\label{MH_pure}
\end{figure}
\begin{figure}[hbt]
	\centering
	\includegraphics[trim=0 0 40 20,clip, width=8.5cm]{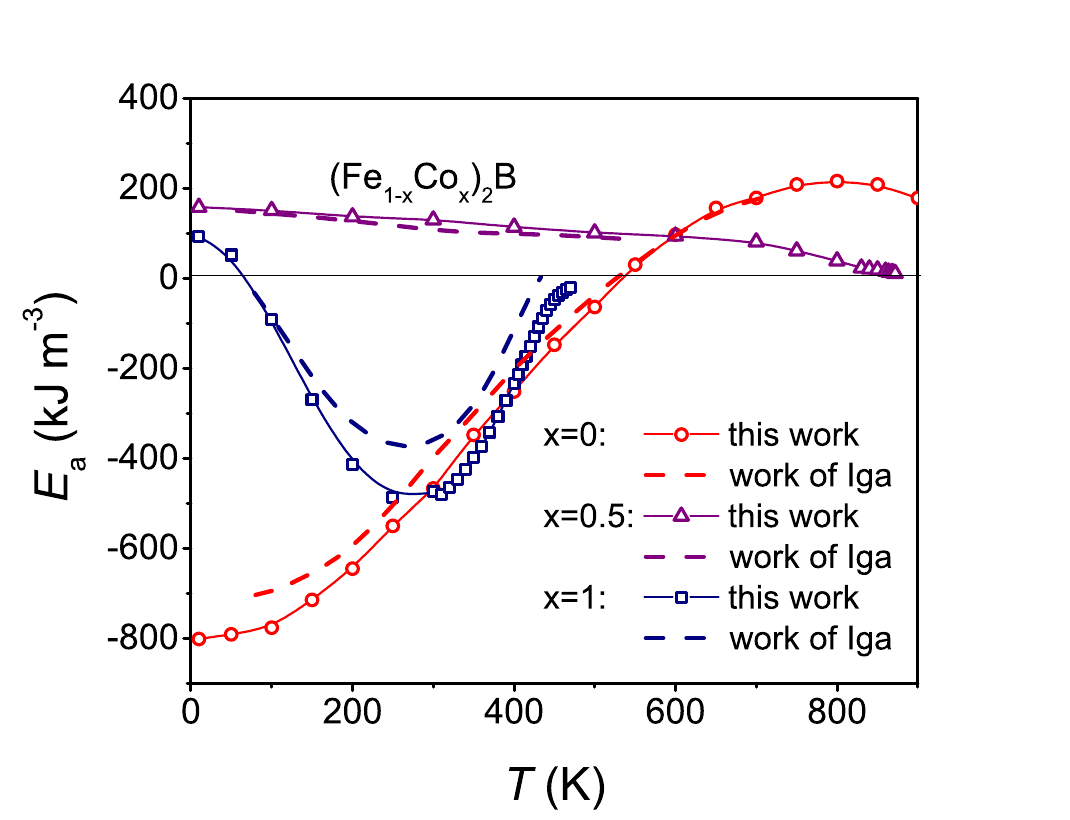}
	\caption{Temperature dependence of the anisotropy energy as determined from the leading anisotropy constant $K_1$ of (Fe$_{1-x}$Co$_{x}$)$_{2}$B. Dashed lines are the values from the work of Iga \cite{iga}}.
	\label{Ea_all}
\end{figure}

Figure \ref{MH_pure} displays the magnetization curves measured along [100] and [001] for Co$_2$B (Fig.~\ref{MH_pure}a) and Fe$_2$B (Fig.~\ref{MH_pure}b) single crystals. All of $M(H)$ curves depicted in Fig.~\ref{MH_pure} are shown after correction on demagnetization factor. The Co$_2$B compound has uniaxial magnetic anisotropy at low temperature (magnetic moment is parallel to the c-axis), however at $T_\mathrm{srt}=72$~K the spin reorientation transition occurs, see Fig.~\ref{Ea_all}. As a result at $T_\mathrm{srt}$ the Co$_2$B single crystal becomes magnetically isotropic and above this temperature, till the Curie temperature, the magnetization vector lies in the (001) plane (easy plane anisotropy). Contrary, the Fe$_2$B single crystal has easy plane anisotropy at temperatures below $T_\mathrm{srt}=593$~K and by passing through the isotropic state at $T_\mathrm{srt}$ becomes magnetically uniaxial.

\begin{figure}[hbt]
	\centering
	\includegraphics[width=8.5cm]{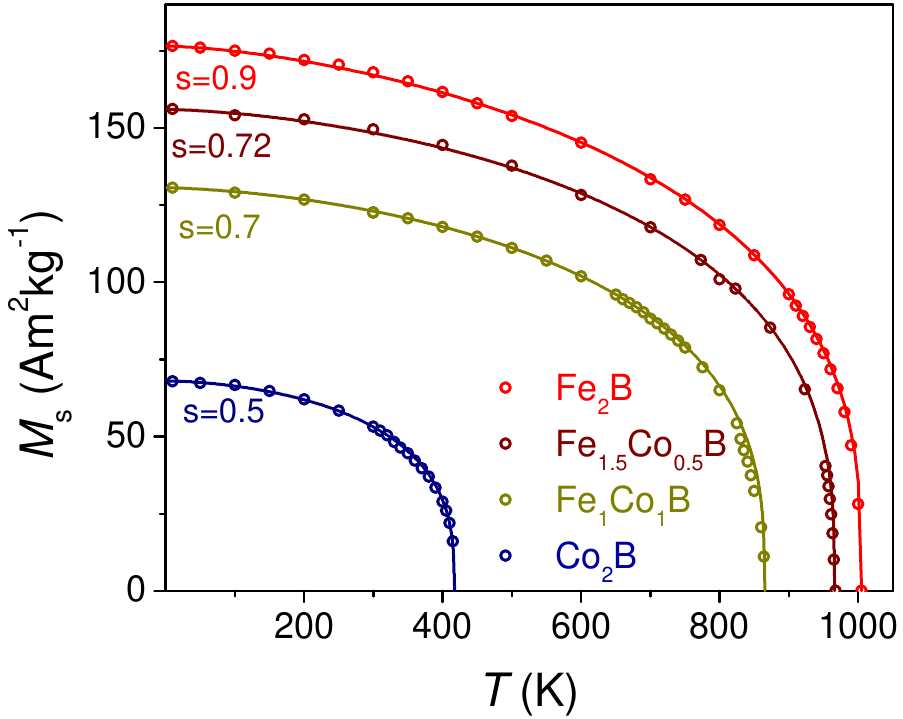}
	\caption{Temperature dependence of the spontaneous magnetization of (Fe$_{1-x}$Co$_{x}$)$_{2}$B.}
	\label{Ms_all}
\end{figure}

Spontaneous magnetizations, $M_s$, determined from the easy axis magnetization curves are plotted versus temperature in Fig.~\ref{Ms_all}. At low temperatures the $M_s$ was determined as the ordinate of the crossing-point of the linearly extrapolated high-field portions of the easy-axis curves (as shown on the 10 K curve in Fig.~\ref{MH_pure}). For a more accurate determination of $M_s$ near the $T_C$ (above 800 K) Belov--Arrott plots \cite{Belov1956,Arrott1957} were used. The continuous line is a fit to the following expression \cite{Kuzmin2014}:
\begin{equation}
	M_s(T)=M_s(0)\left[1-s\left(\frac{T}{T_C}\right)^{\frac{3}{2}}-\left(1-s\right)\left(\frac{T}{T_C}\right)^{\frac{5}{2}}\right]^{\frac{1}{3}}
\end{equation}
where $M_s(0)$ is spontaneous magnetization at 0~K and $T_\text{C}$ is the Curie temperature. Values of the fitting parameter $s$ are shown in Fig.~\ref{Ms_all}.

The anisotropy energy $E_\text{a}$ was determined as the area between the magnetization curves along [100] and [001] taken at the same temperature. No demagnetization correction was necessary since the sample dimensions in both directions had been made practically equal. The so obtained $E_\text{a}$ is presented as a function of temperature in Fig.~\ref{Ea_all}. Also shown in Fig.~\ref{Ea_all} are data from an early work of Iga \cite{iga}. One can appreciate that Iga's values of $K_1$ are noticeably lower than ours. The brevity of Iga's paper \cite{iga} precludes us from making any definite statement about the source of his underestimation; Iga's $K_1$ were deduced from torque curves taken in a field of 1.6~T, which is just above the anisotropy field at low temperatures. In an oblique orientation, such a field is insufficient for making the magnetization vector parallel to the applied field. The remaining misalignment distorts the torque curves and has to be corrected for. Unfortunately, there is no mention of such a correction in Ref.~\cite{iga}. We are led to conclude that (Fe$_{1-x}$Co$_x$)$_2$B are more strongly anisotropic than thought previously.

\section{Density Functional Theory Calculations of $(\text{Fe}_{1-x}\text{Co}_{x})_2\text{B}$ Alloys}\label{sec:dft}


\subsection{Fe$_2$B and Co$_2$B}

Theoretical analysis of \fecob{} alloys starts from the study of the boundary compounds Fe$_2$B and Co$_2$B.
Both these systems crystallize in the tetragonal structure with the space group I4/mcm.
The calculated equilibrium lattice parameters are $a=5.05$~\AA{}, $c=4.24$~\AA{} for Fe$_2$B (cf. experimental values: $a=5.11$~\AA{}, $c=4.25$~\AA{}) and $a=4.97$~\AA{}, $c=4.24$~\AA{} for Co$_2$B (cf. experimental values: $a=5.02$~\AA{}, $c=4.22$~\AA{}).
The optimized Wyckoff positions differ only negligibly from rational numbers (1/6, 2/3 and 1/4).
Thus, the calculations are based on the latter ones (see Tab.~\ref{tab:wyckoff}).
Structural optimizations are performed with the full potential linearized augmented plane wave method (FP-LAPW) implemented in the WIEN2k code~\cite{Blaha01}.
The relativistic effects are included within the scalar relativistic approach with the second variational treatment of spin-orbit coupling.
For the exchange-correlation potential the generalized gradient approximation (GGA) in the Perdew, Burke, Ernzerhof form (PBE)~\cite{Perdew1996} is used. 
Calculations are performed with a plane wave cut-off parameter $RK_\text{max}=8$, total energy convergence criterion $10^{-8}$~Ry and with $20 \times 20 \times 20$ \textbf{k}-points which was tested to give well converged MAE values.

\begin{table}[ht]
\caption{\label{tab:wyckoff} 
Atomic coordinates for Fe$_2$B and Co$_2$B.
}
\begin{tabular}{|l|l|ccc|}
\hline
Atom  	& Site	& $x$ 	& $y$ & $z$ \\
\hline
Fe/Co  	& 8(h)	& 1/6 	& 2/3 & 0   \\
B   	& 4(a) 	& 0 	& 0   & 1/4 \\
\hline
\end{tabular}
\end{table}

The magnetocrystalline anisotropy energy (MAE) is evaluated as the difference between total energies calculated for [100] and [001] quantization axes.
The initial results for MAE are -0.052~meV/f.u. (-0.31~MJ/m$^3$) for Fe$_2$B and -0.168~meV/f.u. (-1.03~MJ/m$^3$) for Co$_2$B.
The experimental values of the anisotropy constant $K_1$ are -0.80~MJ/m$^3$ for Fe$_2$B and +0.10~MJ/m$^3$ for Co$_2$B, while $K_2$ are insignificant\cite{iga}.
Although both theoretical and experimental anisotropy constants indicate the in-plane anisotropy and have the same order of magnitude for Fe$_2$B, for Co$_2$B this comparison appears worse.
A larger discrepancy between theoretical (-1.03~MJ/m$^3$) and experimental (+0.1~MJ/m$^3$) anisotropy constants for Co$_2$B occurs together with another discrepancy between calculated ($1.09\mu_B$) and measured ($0.8\mu_B$~\cite{Cadeville75}) magnetic moments on the Co atom.
This correlation of variances motivates us to study the dependence of MAE on magnetic moment.
Fully relativistic fixed spin moment (FSM) calculations were carried out based on the Full-Potential Local-Orbital Minimum-Basis Scheme (FPLO-14)~\cite{Koepernik99}. 
These calculations were performed with a $20 \times 20 \times 20$ \textbf{k}-mesh, the PBE form~\cite{Perdew1996} of the exchange-correlation potential and convergence criterion $10^{-8}$~Ha.

\begin{figure}[!ht]
\subfloat[MAE$(\mu)$\label{co2b_MAE_vs_fsm}]{%
\includegraphics[width=0.45\textwidth]{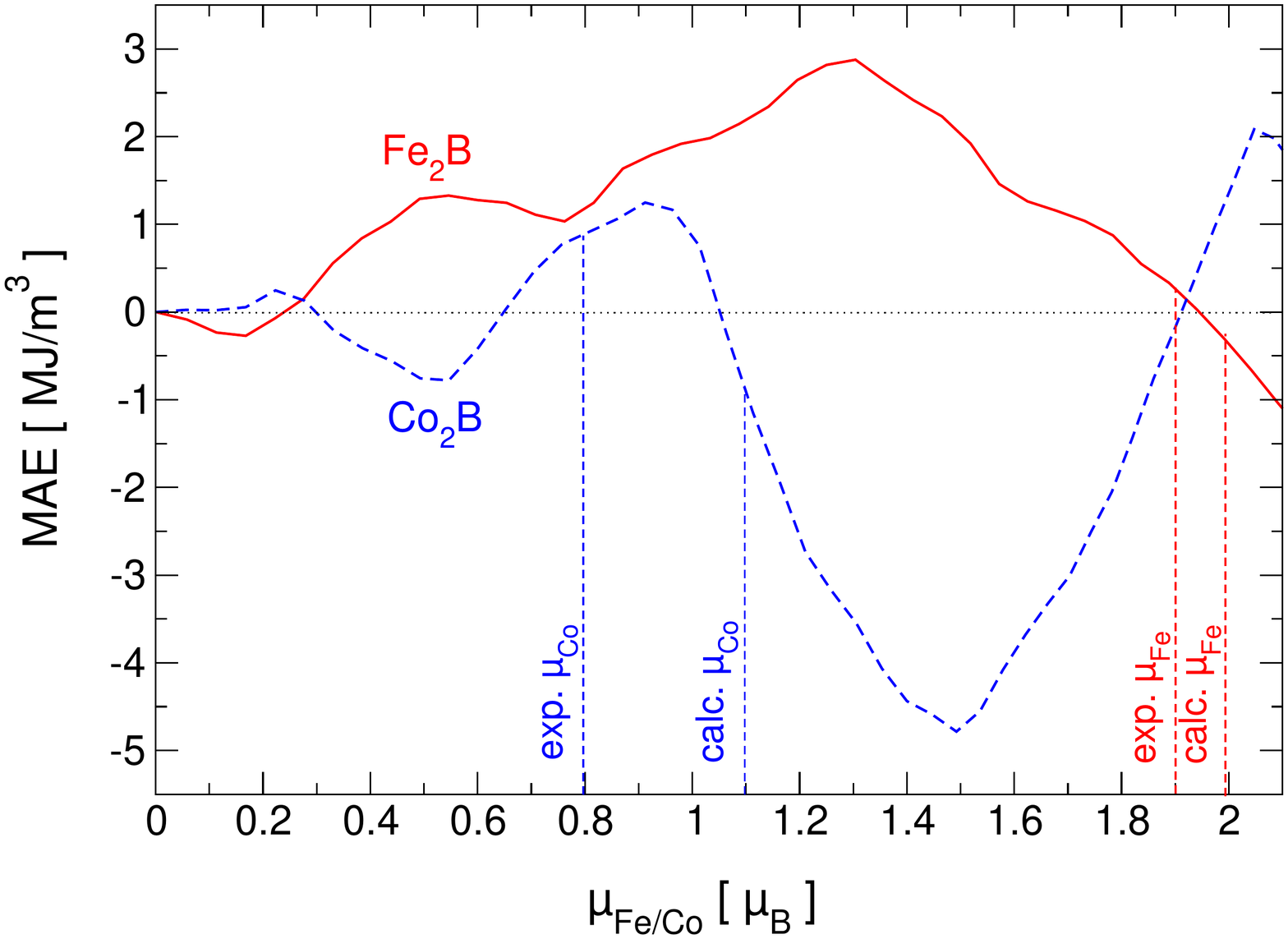}
}
\hfill
\subfloat[MAE$(T)$\label{MAEofT}]{%
\includegraphics[width=0.45\textwidth]{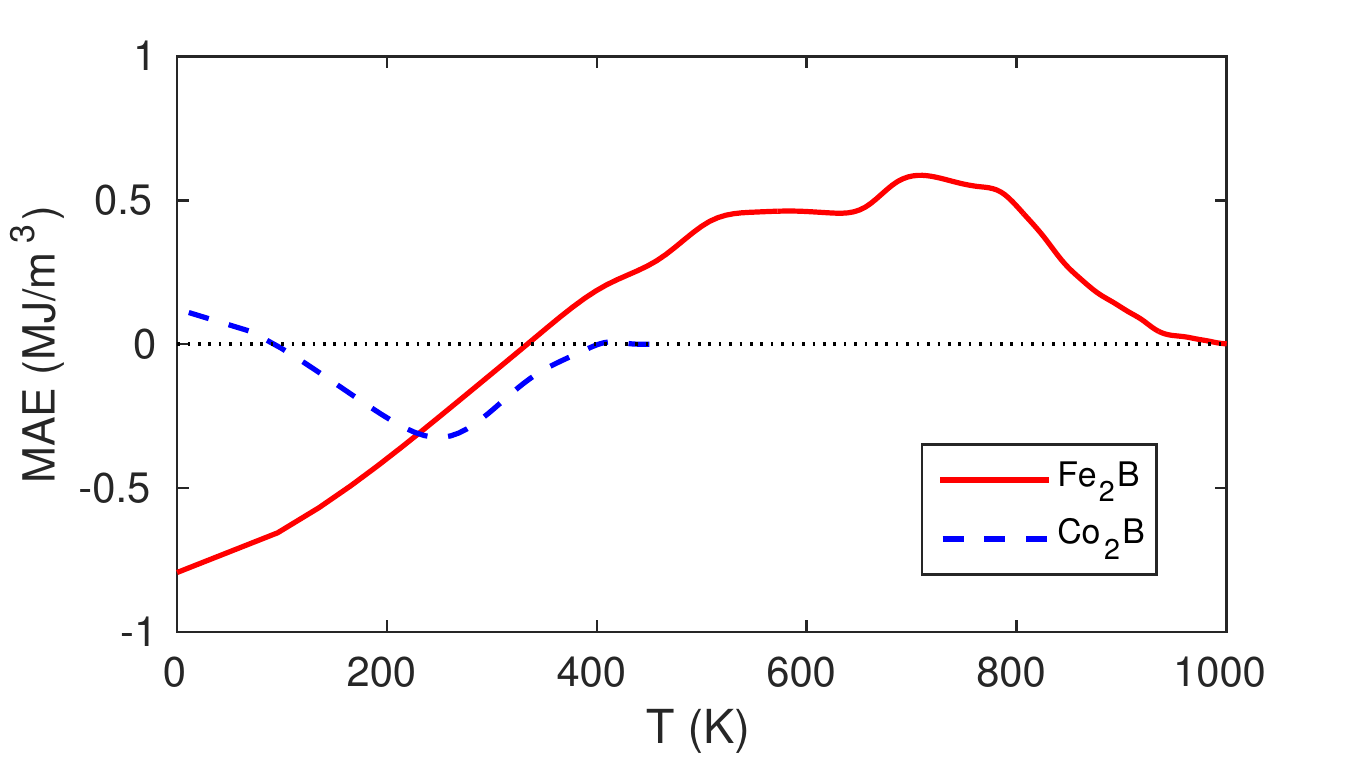}
}
\caption{MAE as function of total magnetic moment on the 3$d$ atom for Fe$_2$B and Co$_2$B as calculated with FSM FPLO in (a). MAE$(T)$ for Fe$_2$B and Co$_2$B as obtained by the scheme described in the text in (b).}
\label{fig:ASAvsFP}
\end{figure}
MAE as a function of magnetic moment (see Fig.~\ref{co2b_MAE_vs_fsm}) indicates, for Co$_2$B, that between the measured ($0.8\mu_B$) and calculated ($1.09\mu_B$) values of magnetic moments, the MAE varies substantially in the range between +1 and -1~MJ/m$^3$.
Furthermore, the slope of MAE as a function of $\mu$ at the point $1.09\mu_B$ is twice bigger for Co$_2$B compared to the slope at $1.98\mu_B$ for Fe$_2$B. 
In summary, FSM calculations show that the MAE of Co$_2$B is a very sensitive function of magnetic moment on Co.
Overestimation of the Co magnetic moment by more than 30\% in the present calculations leads to inaccuracy of the calculated MAE for Co$_2$B.
 
Note that the dependence of MAE on magnetic moment qualitatively can be related with effects of changing temperature and as such it can give clues about the behavior of MAE as a function of temperature.
If we assume that the magnetic moment (magnetization) decreases with $T$, first slowly and then rather rapidly close to $T_C$, we may relate the calculated MAE($\mu$) to experimental MAE($T$).
For Fe$_2$B we see that with decreasing of $\mu$ the MAE starts from a negative value then increases to a positive maximum and after that it decreases towards zero, in a qualitative agreement with the experimental $K_1(T)$ curve.
For Co$_2$B, if we start from the experimental $\mu$ value, we see that MAE starts from a positive value, decreases to negative minimum, and then once again increases, also in qualitative agreement with experimental $K_1(T)$. The analysis above should however be taken with some care since the curves in Fig.~\ref{co2b_MAE_vs_fsm} are from collinear configurations in which each atomic moment decreases, while in the experiments non-collinear configurations certainly also play role. Still, the observation described above gives an indication to the source of the rather complicated temperature dependence of the MAE in this system. It appears that it might be explained by the variation in the MAE due to changes in the band structure as the exchange splitting is reduced with increasing temperature.

In an attempt to quantify the above arguments we present a highly simplified model to map the $\text{MAE}(\mu)$ curves in Fig.~\ref{co2b_MAE_vs_fsm} to temperature. To large extent the temperature dependence of the MAE is expected to come from two sources, namely the reduction in the spin splitting of the bands and the fluctuations in the directions of the moments. The magnetism observed here is expected to be of a highly itinerant character and since the majority spin channel contains unoccupied states, it is itinerant ferromagnetism of the weak kind. Such a system is expected to exhibit a magnetic moment which scales with temperature approximately as\cite{Mohn}
\begin{equation}
m^2 \sim 1-\frac{T^2}{T_\text{C}^2}. \label{eq.mofT}
\end{equation}
Experimental values for Curie temperatures allow an initial mapping of MAE to temperature from the data in Fig.~\ref{co2b_MAE_vs_fsm} via Eq.~\ref{eq.mofT}. In the next step we wish to take into account also the directional fluctuations of the spins. The Callen and Callen model\cite{CallenCallen} predics that such fluctuations result in a $\left< m(T)/m(0)\right>^3$ behaviour in the MAE. A classical mean field model with a Langevin function is used to enforce such a scaling. Finally, we choose the the zero-temperature reference values for the moments in order to obtain MAE$(T=0)$ values closer to experiment and hence a quantitatively somewhat more accurate model than that obtained by using the experimentally or computationally obtained moments at low temperature. In doing so the zero-temperature moments are chosen to $m_\text{Co}(T=0) = 0.7\mu_\text{B}$ and $m_\text{Fe}(T=0) = 2.1\mu_\text{B}$. The resulting $\text{MAE}(T)$ curves for Fe$_2$B and Co$_2$B are included in Fig.~\ref{MAEofT}. It appears that this crude and simplified model at least captures the main features of the experimental $\text{MAE}(T)$ curves when comparing to Fig.~\ref{Ea_all}. This further supports the idea that much of the interesting non-monotonous behaviour in the $\text{MAE}(T)$ is due to variations in the band structure as temperature reduces the exchange splitting.

\subsection{Virtual Crystal Approximation}

To study the whole range of compositions between Fe$_2$B and Co$_2$B the Virtual Crystal Approximation (VCA) method was used.
VCA imitates a random on-site occupation of two types of atoms by using a virtual atom with an averaged value of the nuclear charge $Z_{\text{VCA}}$.
In the case of the \fecob{} alloys it is $Z_{\text{VCA}}~=~(1-x) \cdot Z_{\text{Fe}} + x \cdot Z_{\text{Co}} = 26 + x$.
The number of valence electrons on the VCA site is also increased correspondingly. 

\begin{figure}[htb]
\centering
\includegraphics[width=8.6cm]{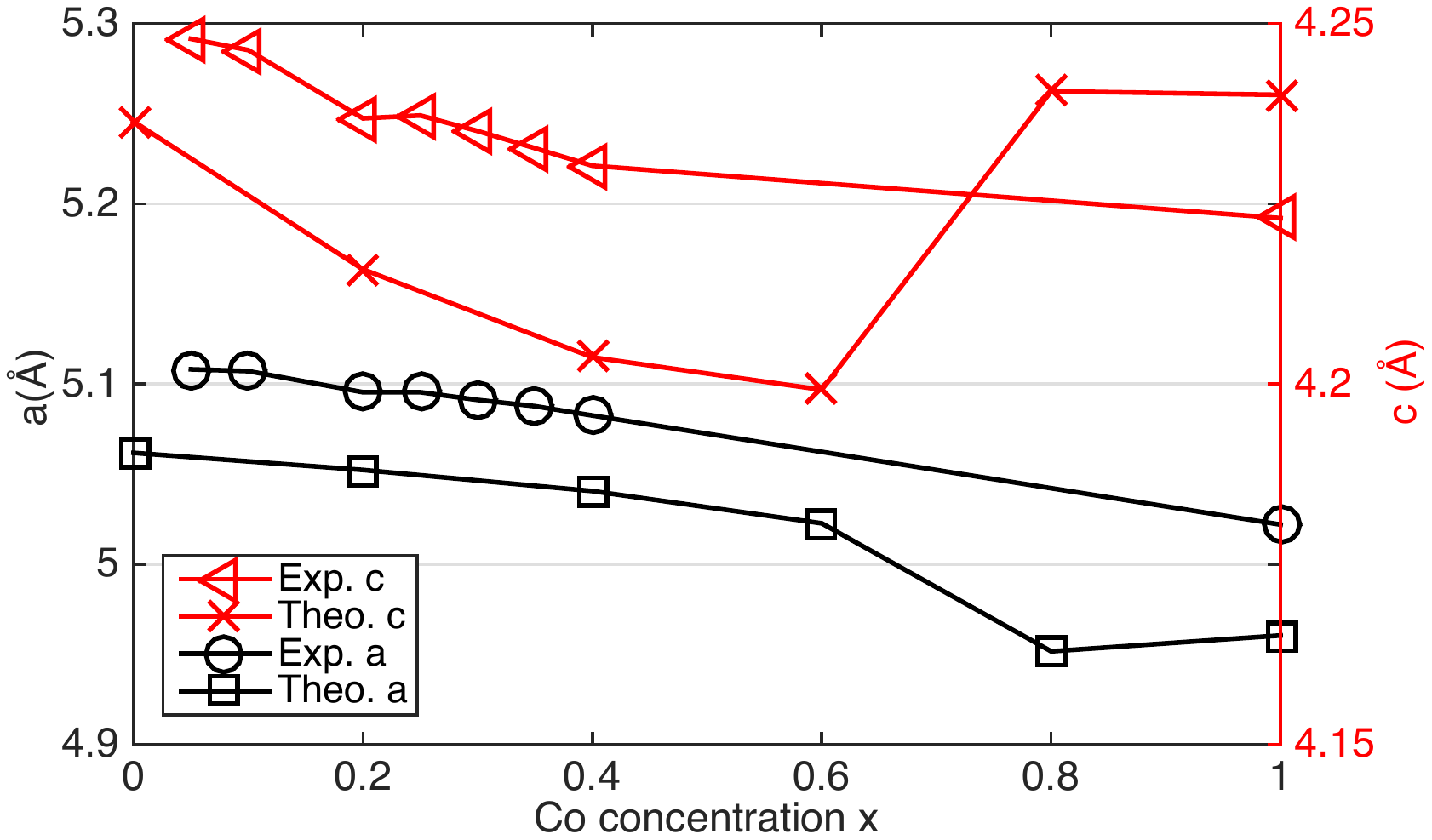}
\caption{Lattice parameters as functions of $x$ in \fecob{}, from experiment and calculated with WIEN2k in the GGA, treating disorder by the VCA.}
\label{lattparfig}
\end{figure}

In the first step of investigating the alloy compound, the lattice parameters are studied as a function of alloy concentration using WIEN2k within the GGA.
Calculations were performed with and without the inclusion of SOC which, however, had a negligible effect on lattice parameters. 
Fig.~\ref{lattparfig} shows experimentally measured and theoretically calculated lattice parameters, $a$ and $c$, as functions of $x$ in \fecob{}.

Overall, the discrepancy between theory and experiment is small and in general the variation in lattice parameters are small, in the order of 1-2\% over the whole range. 
Calculations were also performed within the local density approximation (LDA)\cite{lsda} with similar results for $a$ and $c$ but somewhat greater disagreement with experiment (not shown).
Further results presented in this section are obtained with structures from linear interpolation between theoretically obtained lattice parameters of the end compounds. For a few alloy concentrations the MAE obtained by using the computationally optimized lattice parameters was compared to that obtained by using the lattice parameters from linear interpolation and the difference was found to be negligible.
It was also investigated how internal atomic positions vary with $x$ and the variation was found to be minute, within numerical and experimental uncertainties, whereby fixed values were assumed as presented in Tab.~\ref{tab:wyckoff}.

\begin{figure}[htb]
\centering
\includegraphics[width=8.6cm]{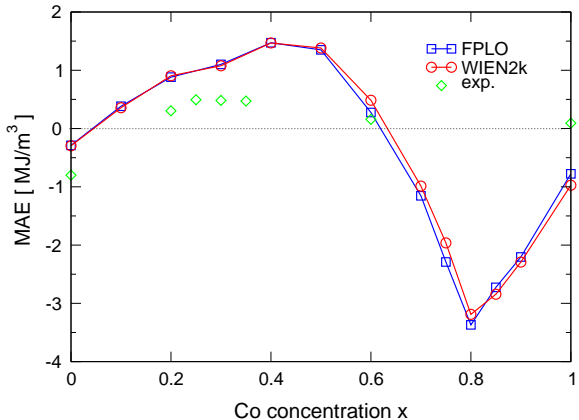}
\caption{MAE as a function of $x$ in \fecob{} calculated with WIEN2k and FPLO, in both cases treating disorder by the VCA.}
\label{feco2b_mae_vs_x_fplo_wien2k}
\end{figure}

Fig.~\ref{feco2b_mae_vs_x_fplo_wien2k} reveals an excellent agreement between MAE's as functions of $x$ calculated with the WIEN2k and FPLO codes. 
Such consistency reflects a good convergence and high numerical accuracy of these results. 
The function MAE$(x)$ is in good qualitative agreement with measured $K_1(x)$ at low temperatures\cite{iga}.
Both the uniaxial anisotropy maximum, closer to Fe-rich side, and the in-plane anisotropy minimum, on Co-rich side, are reproduced.
The highest uniaxial anisotropy, with MAE~=~1.5 MJ/m$^3$, occurs for the (Fe$_{0.6}$Co$_{0.4}$)$_2$B alloy. 
%
%
Quantitative overestimation of the MAE for the whole concentration range brings reminiscence to the case of tetragonally strained Fe/Co alloys, where VCA treatment of substitutional disorder\cite{Burkert2004} also led to an overestimation of MAE compared to experiments\cite{Andersson2006} or more sophisticated treatments of disorder\cite{Neise2011,Turek2012}.
Therefore, in the next subsection we will also present results based on a more advanced description of disorder, the coherent potential approximation (CPA).

\begin{figure}[htb]
\centering
\includegraphics[width=8.6cm]{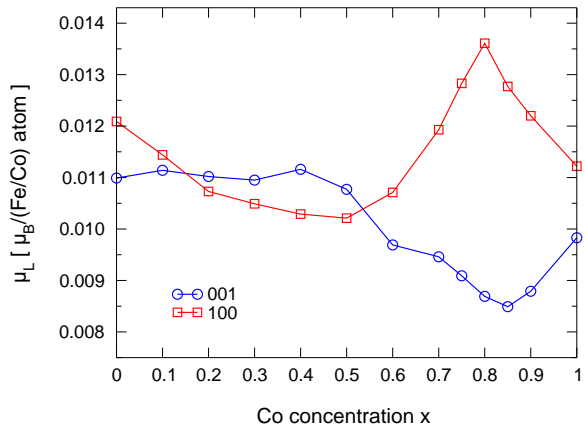}
\includegraphics[width=8.6cm]{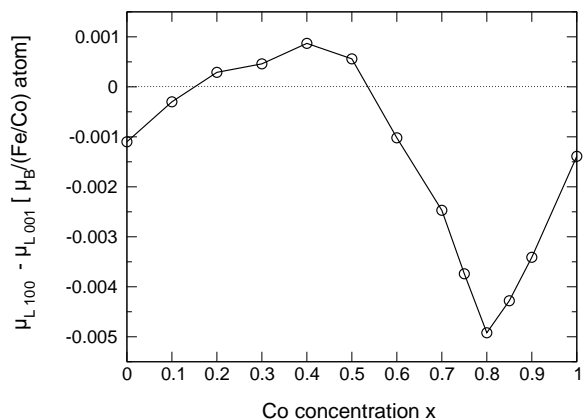}
\caption{The orbital magnetic moment $\mu_L$ and difference of $\mu_L$'s for [100] and [001] quantization axis as functions of $x$ in \fecob{}, calculated with FPLO treating disorder by the VCA.}
\label{feco2b_orb_mm}
\end{figure}

We continue first with our analysis of the VCA results by exploring the relation of the MAE to the orbital angular momentum anisotropy. Calculated orbital magnetic moments $\mu_L$ vary between 0.008 and 0.014~$\mu_B$ per Fe/Co atom.
As seen from Figs.~\ref{feco2b_mae_vs_x_fplo_wien2k} and \ref{feco2b_orb_mm} the MAE closely follows the orbital moments anisotropy, as suggested by Bruno's formula \cite{Bruno1989}, although with a slight shift towards the positive values.

\begin{figure}[htb]
\centering
\includegraphics[width=8.6cm]{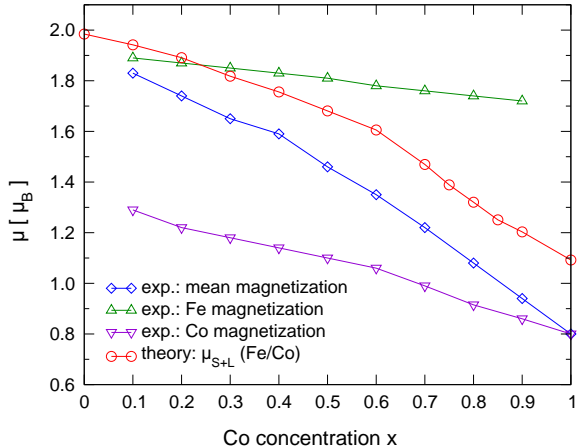} 
\caption{Total and species-resolved magnetic moments as functions of $x$ in \fecob{}. Experimental values from Ref.~\cite{Cadeville75}, theoretical moments calculated with FPLO treating disorder by the VCA.}
\label{feco2b_mm}
\end{figure}

Total magnetic moment $\mu_{S+L}$ per (Fe/Co) virtual atom calculated as a function of $x$ in \fecob{} is compared with experimental values of mean magnetization\cite{Cadeville75} in Fig.~\ref{feco2b_mm}. 
Discrepancy between experiment and theory is equal to 0.1~$\mu_B$/atom for Fe$_2$B and 0.3~$\mu_B$/atom for Co$_2$B.
For intermediate concentrations discrepancies scale almost linearly with $x$.

\begin{figure}[htb]
\centering
\includegraphics[width=8.6cm]{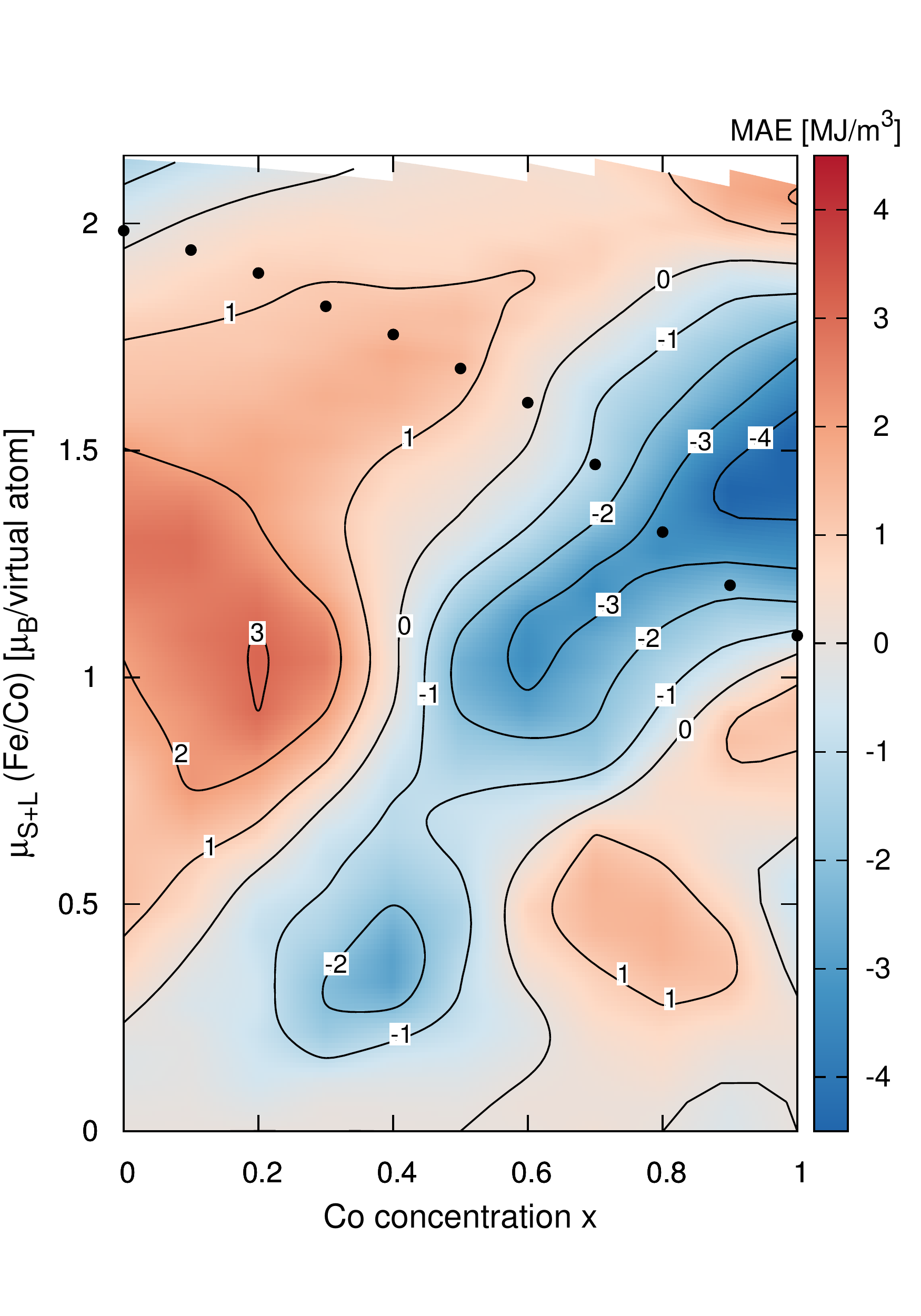}
\caption{Fully relativistic FPLO calculations of MAE as a function of $x$ and total magnetic moment ($\mu_{S+L}$) on 3d atom for \fecob{}.
Disorder was treated by the VCA and $\mu_{S+L}$ were stabilized with fixed spin moment FSM approach. Equilibrium $\mu_{S+L}$'s (see Fig.~\ref{feco2b_mm}) are denoted by black dots.}
\label{feco2b_color_map}
\end{figure}

In Fig.~\ref{feco2b_color_map} the FSM and VCA dependencies of MAE are presented as a 2D color map in superposition with the theoretical equilibrium magnetic moments $\mu_{S+L}$'s denoted by black dots.
The MAE landscape reveals that by going from magnetic moment of pure Fe$_2$B towards the much lower magnetic moment of Co$_2$B the MAE path passes across a broad range of positive values and a narrower and steeper valley of negative values.
Particularly, the final value of MAE on the Co$_2$B side depends sensitively on the Co magnetic moment.
Clearly, on the Co-rich side the overestimation of Co moment affects substantially the shape of MAE curve.
Focusing on the uniaxial anisotropy, the highest MAE value along the equilibrium path is reached for the (Fe$_{0.6}$Co$_{0.4}$)$_2$B alloy.

We note that on the calculated MAE landscape, there is a region with about three times higher values of MAE for $\mu$ about 1.25~$\mu_B$/atom on the Fe-rich side.
To approach this region, starting, e.g., from the (Fe$_{0.6}$Co$_{0.4}$)$_2$B composition, both the magnetic moment and the $d$-electrons number (proportional to $x$) should be reduced.
The desired magnetic moment reduction from 1.75 to around 1.25 $\mu_B$/atom is around 25\%.
This can be obtained by alloying Fe$_{0.6}$Co$_{0.4}$ with 25\% of a suitable non-magnetic element.
The non-magnetic alloying should at the same time decrease $x$ (which may be understood as the number of $d$ electrons beyond those of Fe) by about 0.2.
Such decrease can be obtained by alloying Fe$_{0.6}$Co$_{0.4}$ with elements having less of $d$ electrons, i.e., the elements from columns of the periodic table preceding the column with Fe.
Among $3d$ elements which fulfill the latter condition are Cr or Mn, but these often carry rather large magnetic moments, therefore they would not fulfill the first condition -- reduction of average moment per atom. From $4d$ and $5d$ elements, possible candidates are Mo and Tc, or W and Re, respectively. Doping with heavier transition metal elements will be explored in Sec.~\ref{sec:imp}.


\subsection{Coherent Potential Approximation}\label{CPAsec}

As seen in the previous section, the VCA yields a correct qualitative agreement for MAE as a function of $x$ in \fecob{} when comparing with experiments. However, as previously discussed, it quantitatively overestimates the MAE. Hence, calculations have also been performed using the coherent potential approximation (CPA) to treat the alloying. These calculations were carried out with the SPRKKR\cite{sprkkr, Ebert2012} method and GGA\cite{Perdew1996} for the exchange-correlation potential. The MAE was evaluated by the torque method\cite{Wang1996} and as much as 160000 $\hat{k}$-vectors were used for the numerical integration over the full Brillouin zone in order to obtain MAE values with numerical accuracy within a few percent. Calculations were done using the lowest temperature, i.e. 77~K, experimental lattice parameters reported by Iga\cite{iga}.

\begin{figure}[hbt]
\centering
\includegraphics[width=8.6cm]{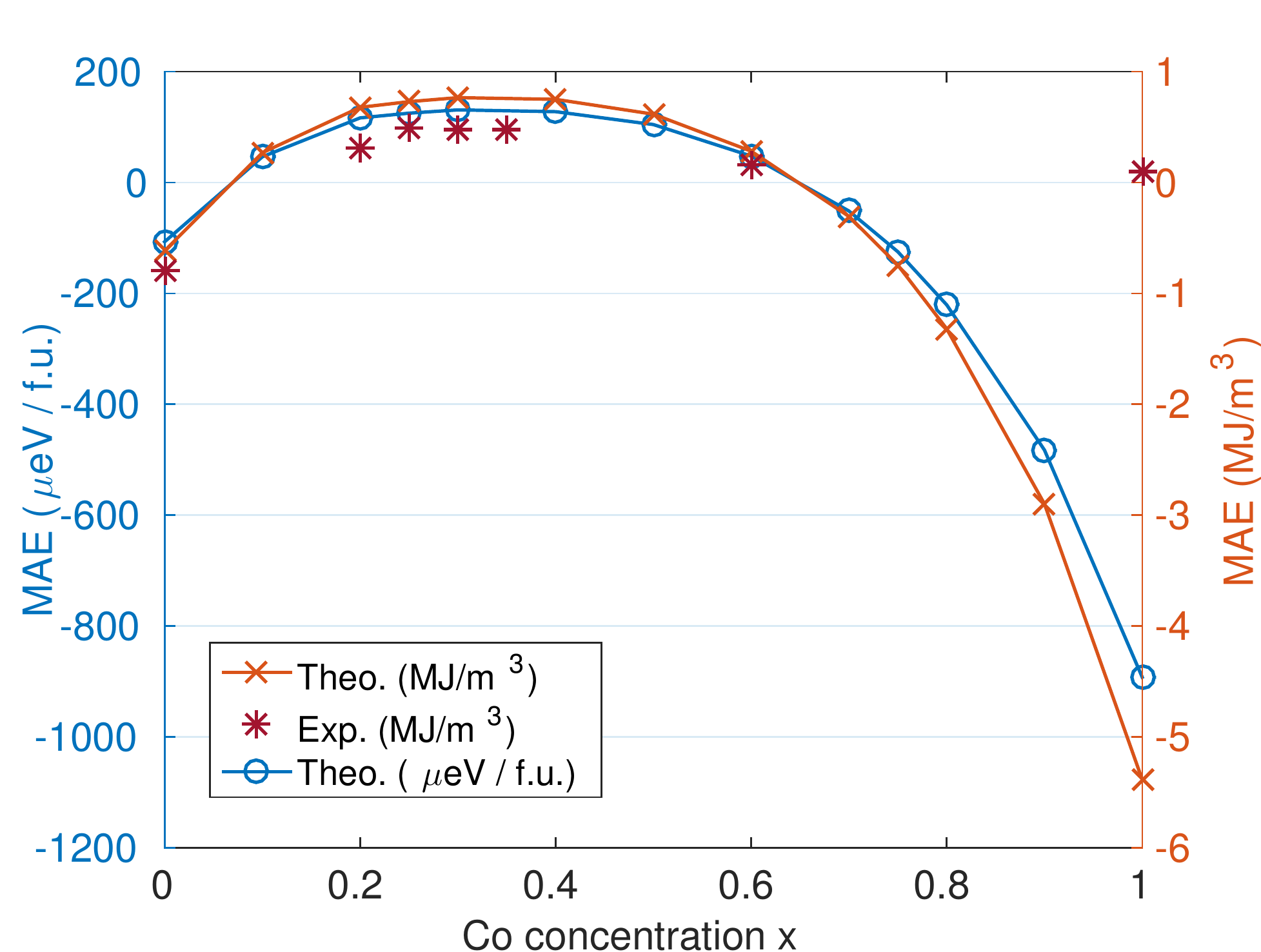}
\includegraphics[width=8.6cm]{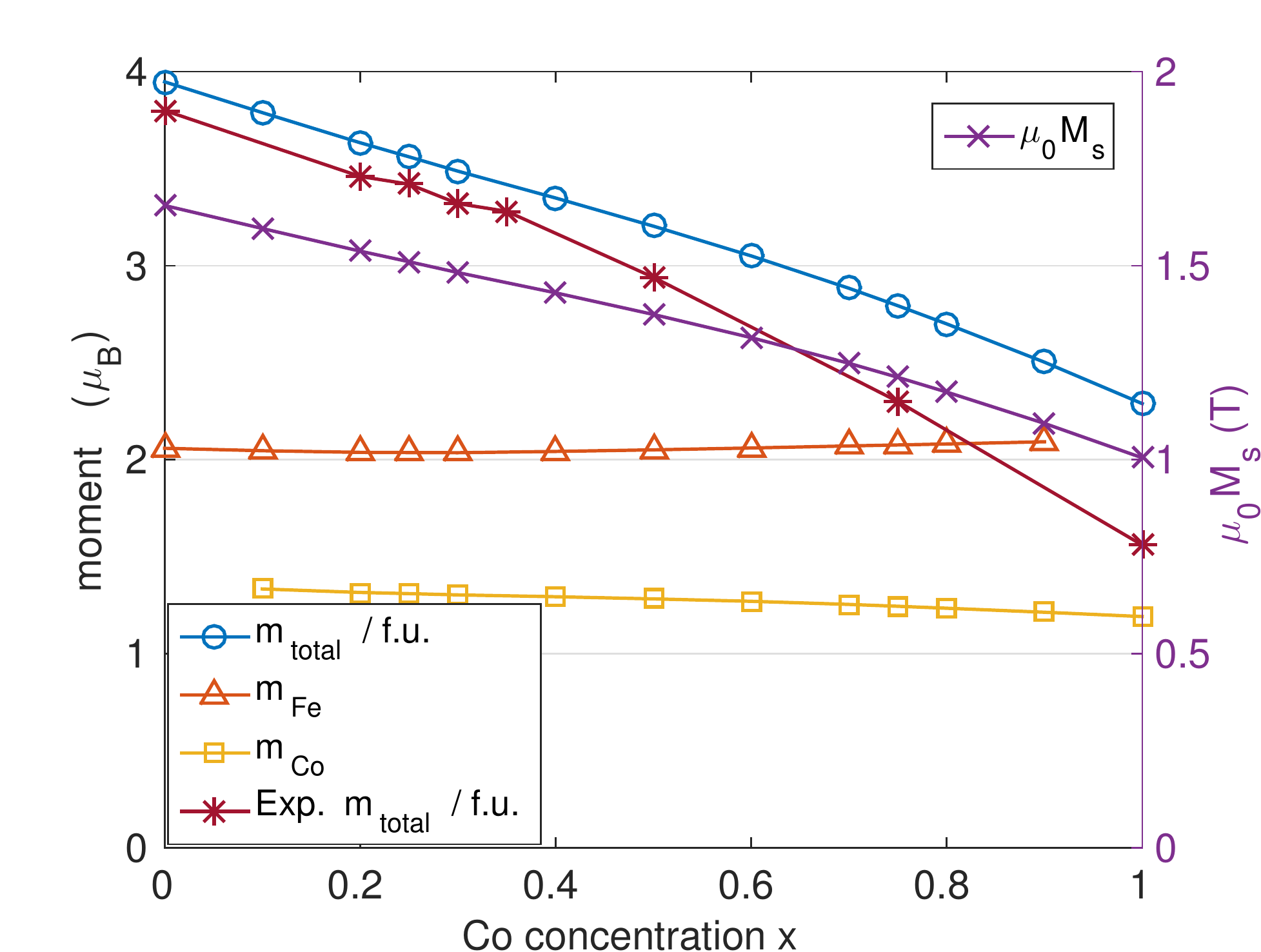}
\caption{MAE and saturation magnetization as functions of $x$ in \fecob{} as calculated with SPRKKR treating disorder by the CPA.}
\label{CPAofx}
\end{figure}

Fig.~\ref{CPAofx} shows the MAE$(x)$ and saturation magnetization, $M_\text{s}(x)$, as result of these calculations. As expected, application of CPA instead of VCA removes the problem of significantly overestimating the MAE of the alloy, at least on the Fe-rich side of the diagram. Instead the qualitative behavior appears to be inconsistent with the experiments of Iga\cite{iga}, as well as the experimental data provided in this paper, on the Co-rich side of the diagrams. This failure is most likely due to the atomic sphere approximation (ASA) which approximates the potential as being spherically symmetric around each atom, in contrast to the VCA calculations presented in the previous section, which included full potential (FP) effects. This is in contrast with a recent publication\cite{belashchenko}, where authors suggest that the qualitatively wrong behavior in the MAE predicted for the Co-rich side is a failure of exchange-correlation functional itself, yielding too high magnetization. Nevertheless, the CPA description in ASA coincides with experiments on the Fe-rich side. In particular the region around $0.2 \leq x \leq 0.4$, with rather large uniaxial MAE, which is the most interesting region from a practical perspective, is correctly described. The maximum MAE is found for $x=0.3$ where $\text{MAE} = 131~\mu\text{eV/f.u.} = 0.77~\text{MJ/m}^3$.

In order to further investigate the importance of full potential (FP) effects, calculations using both ASA and FP in SPRKKR were performed for Fe$_2$B and Co$_2$B. The band structures around the Fermi energy resulting from these calculations are shown in Fig.~\ref{fig:ASAvsFP}. In the case of Fe$_2$B, there are quantitative changes in the bands but qualitatively they look similar. The MAE of the system changes from $-0.11~\text{meV/f.u.} = -0.64~\text{MJ/m}^3$ in ASA to $-0.22~\text{meV/f.u.}=-1.28~\text{MJ/m}^3$  in FP, which can be compared to the experimental value of $-0.80~\text{MJ/m}^3$. This variation in MAE is certainly of numerical significance but still the numbers qualitatively agree on the sign and order of magnitude and both are reasonably close to experiment. 

\begin{figure}[!ht]
    \subfloat[Fe$_2$B\label{Fe2B_bands}]{%
      \includegraphics[width=0.45\textwidth]{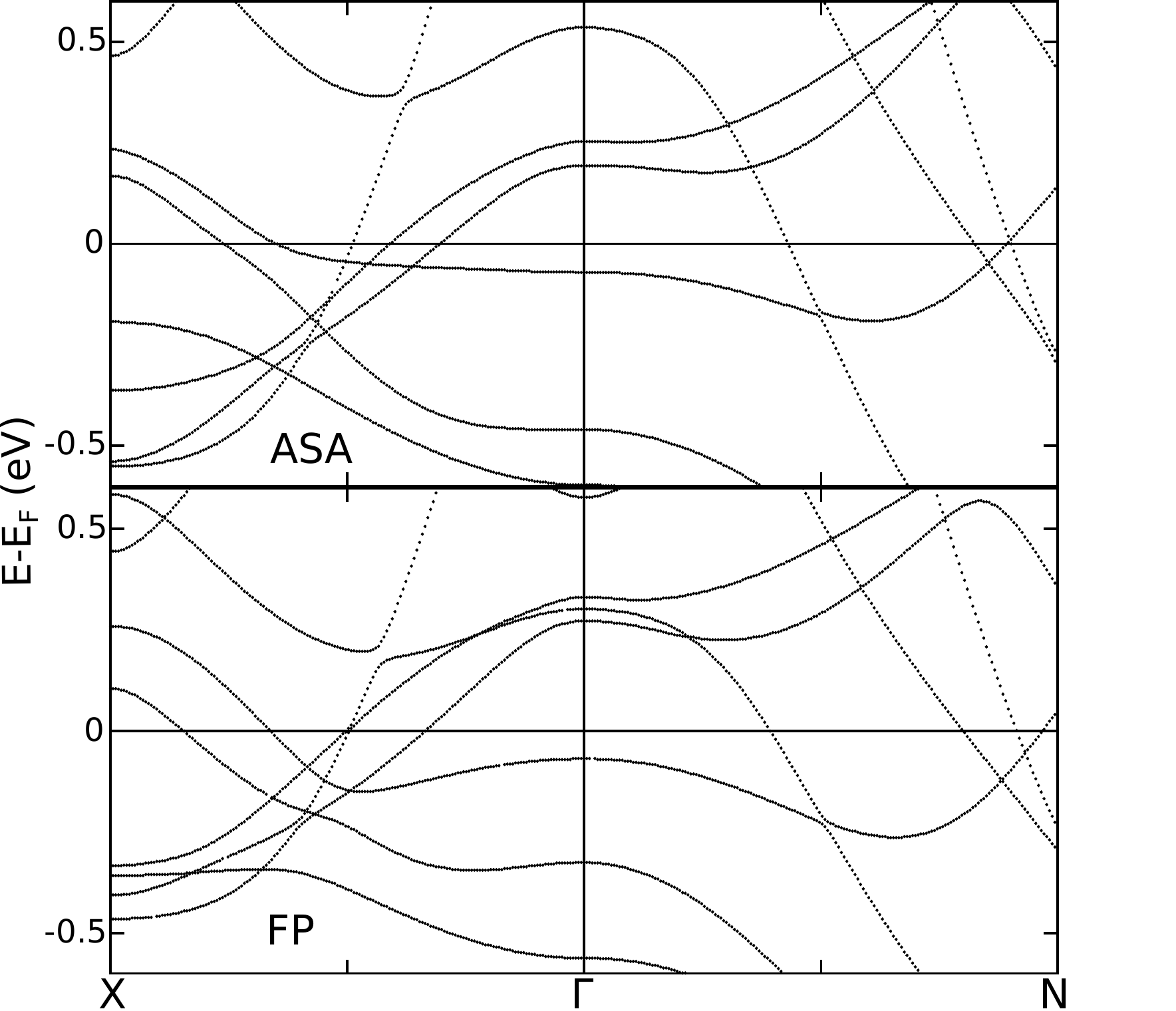}
    }
    \hfill
    \subfloat[Co$_2$B\label{Co2B_bands}]{%
      \includegraphics[width=0.45\textwidth]{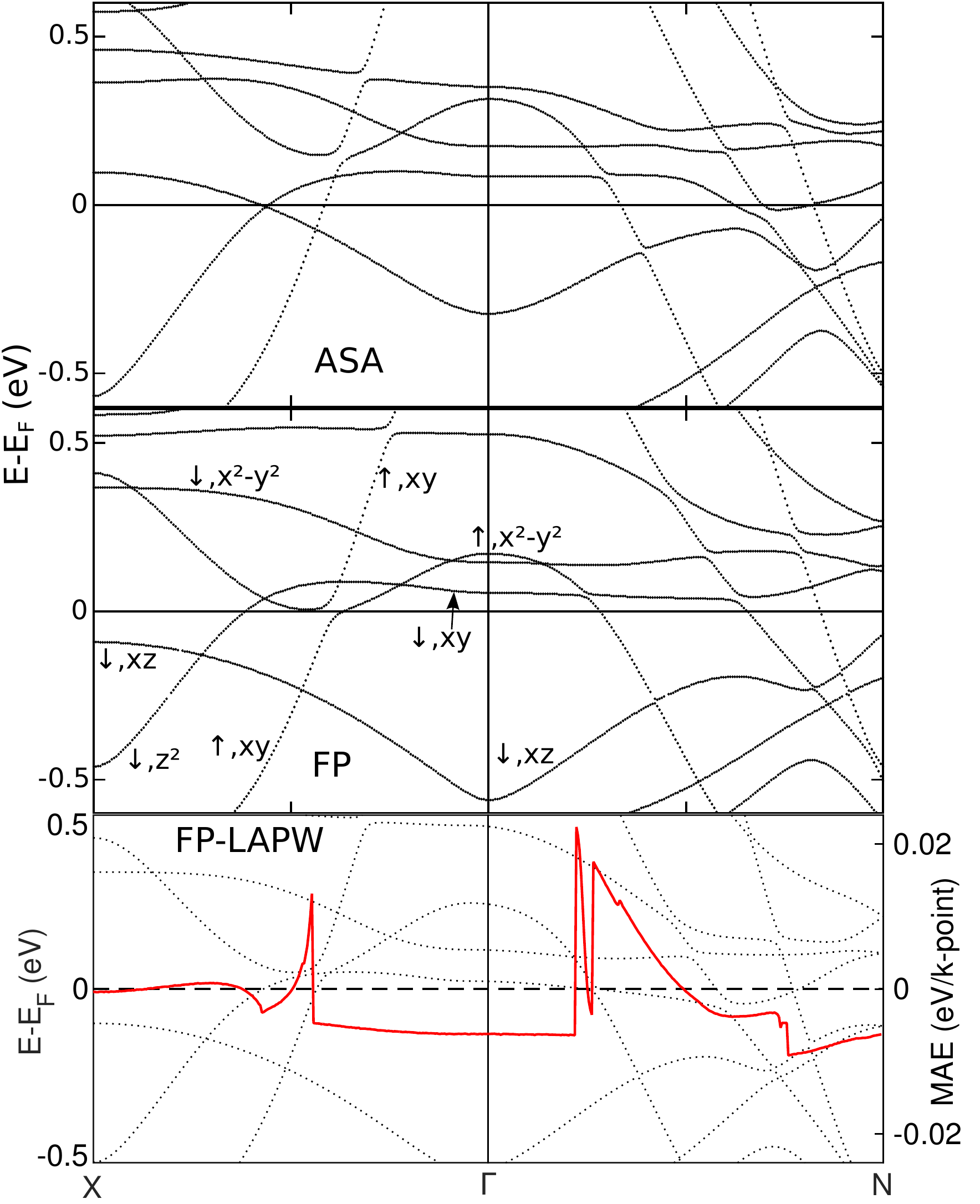}
    }
    \caption{Band structure of Fe$_2$B and Co$_2$B obtained in SPRKKR-ASA and SPRKKR-FP and for Co$_2$B also that obtained with FP-LAPW calculations, in the lower panel,together with the MAE contribution of each {\bf k}-point (solid red line) as obtained by the force theorem. The orbital and spin characters of some bands are indicated with majority and minority spin denoted by $\uparrow$ and $\downarrow$, respectively.}
    \label{fig:ASAvsFP}
  \end{figure}

For the case of Co$_2$B, the MAE changes from $-0.89~\text{meV/f.u.}=-5.4~\text{MJ/m}^3$ to $-0.37~\text{meV/f.u.}=-2.2~\text{MJ/m}^3$ upon inclusion of FP in the calculations. The later value is significantly closer to the experimental low temperature values, as well as to the results of FP calculations presented in previous sections of this paper and it could potentially allow for reproducing the correct qualitative behavior for the Co-rich alloys. In the band structure one can observe certain qualitative changes which can be of relevance with respect to the MAE. In particular around the high symmetry point X, a relatively flat band which is unoccupied in ASA becomes occupied in FP which is relevant to the MAE since bands crossing the Fermi energy tend to dramatically affect the MAE\cite{Wu}. However, to allow for further analysis of the importance of various bands to the MAE, we have used scalar relativistic, spin polarized calculations, neglecting spin-orbit interaction, carried out in WIEN2k to identify the orbital and spin characters of some bands which are marked in Fig.~\ref{Co2B_bands}. This information can be used in analyzing how the coupling between the bands contribute to the MAE by considering the SOC in second order perturbation theory\cite{Bruno1989}, where the only non-zero contribution comes from coupling between occupied and unoccupied states, and by looking at the spin-orbit matrix elements which are tabulated e.g. in Ref.~\cite{Abate1965}. In the bottom part of Fig.~\ref{Co2B_bands}, the WIEN2k FP-LAPW bands, which are very similar to the SPRKKR-FP bands, are also shown for comparison together with the MAE contribution per $\mathbf{k}$-point as obtained by the magnetic force theorem\cite{Liechtenstein, Wu}. In the ASA case, the highest occupied band at X shows mainly minority spin $d_{z^2}$ character while the lowest unoccupied is minority spin $d_{xz}$. These states do not couple via spin-orbit interaction whereby a weak contribution to the MAE is expected in this region. As the minority spin $d_{xz}$ band becomes occupied in FP, it will couple to both of the unoccupied states with magnetic quantum number $m=\pm2$ which have opposite spin character and cause MAE contribution of opposite signs resulting in cancellation and again a weak effect on the MAE as also seen explicitly in the bottom panel of Fig.~\ref{Co2B_bands}. An important negative MAE contribution instead appears to come from a region around $\Gamma$ and the coupling of the occupied minority spin $d_{xz}$ band with the flat unoccupied minority spin $d_{xy}$ band (in general, coupling between states with the same spin but different magnetic quantum numbers contribute negatively to the MAE\cite{Wu, Abate1965}). These two bands are closer to each other in the ASA case allowing for this negative region to be overestimated and contribute to the too large negative MAE obtained for Co$_2$B in ASA calculations. Typically, flat bands such as that just above the Fermi energy at $\Gamma$ can contribute strongly to the MAE and small shifts in such bands will be essential and might partially explain the sensitivity in the MAE of Co$_2$B to various parameters.
Including FP effects, the MAE is still more strongly negative both than the experimental value, which is slightly positive at low temperatures, and that from previous FP (WIEN2k and FPLO) calculations, which yield a negative value with approximately half the magnitude. However, the previous calculations were performed using computationally optimized lattice parameters while the SPRKKR calculations used slightly larger experimental lattice parameters. Calculating the MAE for Co$_2$B with the computationally optimized lattice parameters presented in previous sections of this paper in SPRKKR-FP yields $\text{MAE} = -0.19~\text{meV/f.u.}=-1.2~\text{MJ/m}^3$, in good agreement with the FPLO calculations ($-0.17~\text{meV/f.u.}=-1.03~\text{MJ/m}^3$). Another indication that FP effects are of great significance for Co$_2$B comes from the density of states at the Fermi energy. For Co$_2$B this quantity changes significantly from $8.1$ to $5.9~\text{states/eV}$ upon inclusion of FP effects, while for Fe$_2$B the change is from $4.4$ to $4.3~\text{states/eV}$. Moreover, the relatively large density of states at the Fermi energy for Co$_2$B could partially explain why the MAE of this compound appears more sensitive to various parameters since, with many states near the Fermi energy, a precise description of these states will be essential for a correct MAE.  In conclusion it is clear that the MAE, which is in general a challenging quantity to obtain from first principles, is for this system, especially on the Co-rich side, particularly sensitive to various parameters and difficult to describe correctly. However, the Fe-rich side of the compound appears to be well described using ASA and CPA, which will hence be used in Sec.~\ref{sec:imp}. Since the full potential VCA calculations overestimate the MAE but produce the qualitatively correct shape of the $\text{MAE}(x)$ curve, while the ASA CPA calculations yield a qualitatively incorrect curve on the Co-rich side but the correct behaviour on the Fe-rich side, one is led to believe that a combination of FP and CPA might produce an MAE curve which is both qualitatively and quantitatively in reasonable agreement with experiment for all alloy concentrations. 

\begin{figure*}[hbt]
\centering
    \subfloat[$x=0.1$\label{x10bsf}]{%
      \includegraphics[trim = 1mm 1mm 10mm 5mm, clip, height=4.1cm]{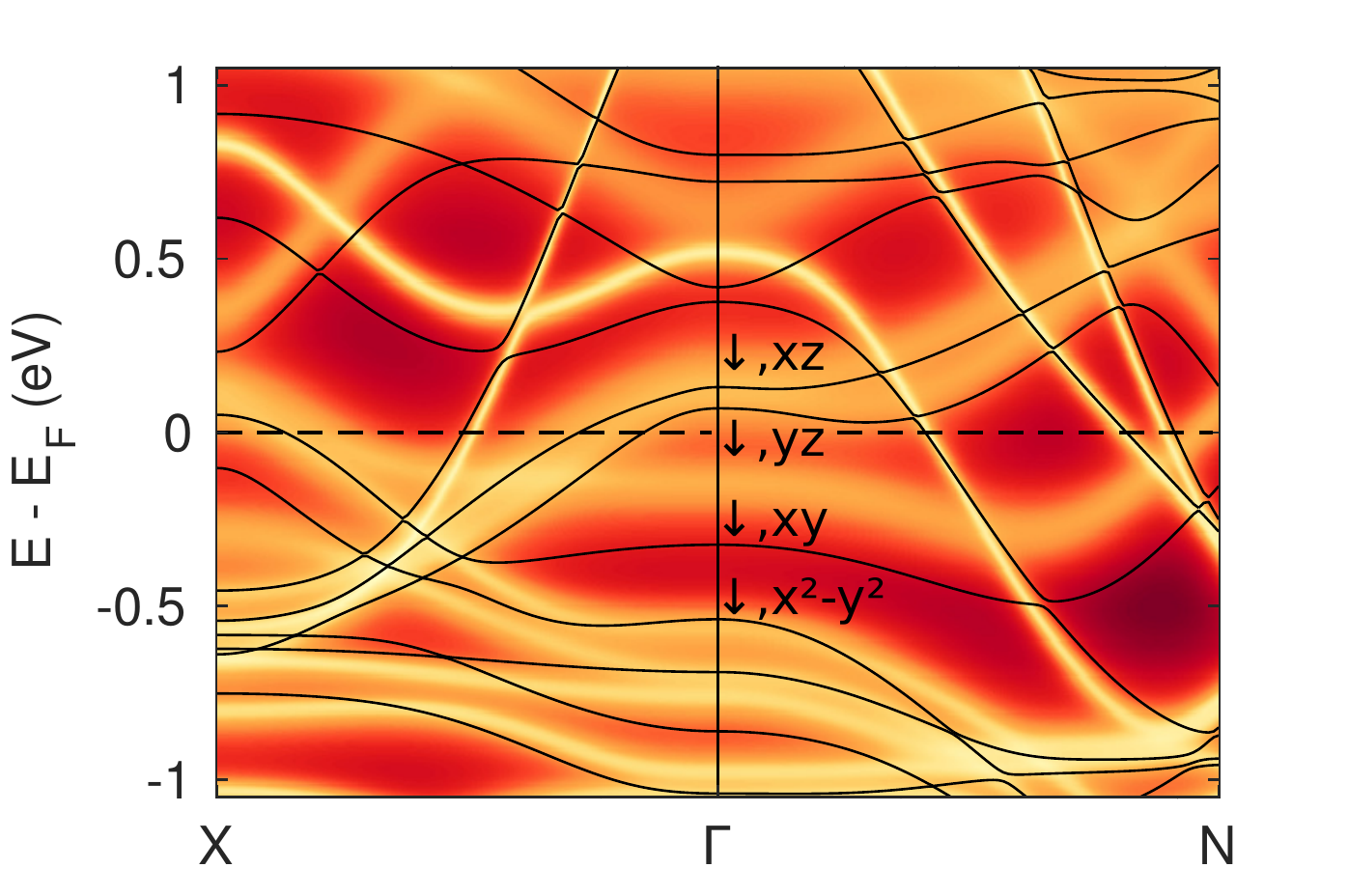}
    } 
    \subfloat[$x=0.2$\label{x20bsf}]{%
      \includegraphics[trim = 10mm 1mm 10mm 5mm, clip, height=4.1cm]{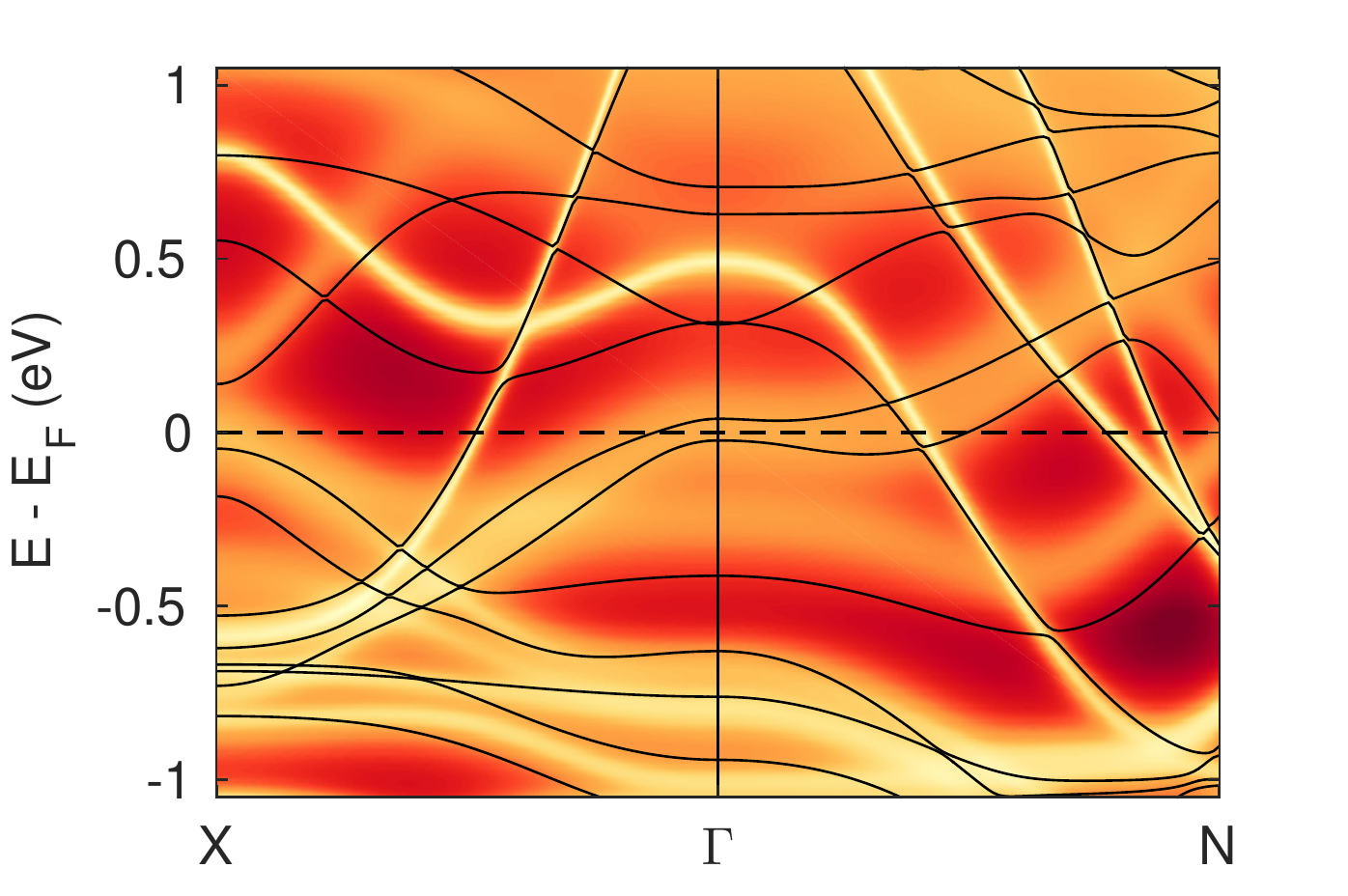}
    } 
    \subfloat[$x=0.3$\label{x30bsf}]{%
      \includegraphics[trim = 10mm 1mm 10mm 5mm, clip, height=4.1cm]{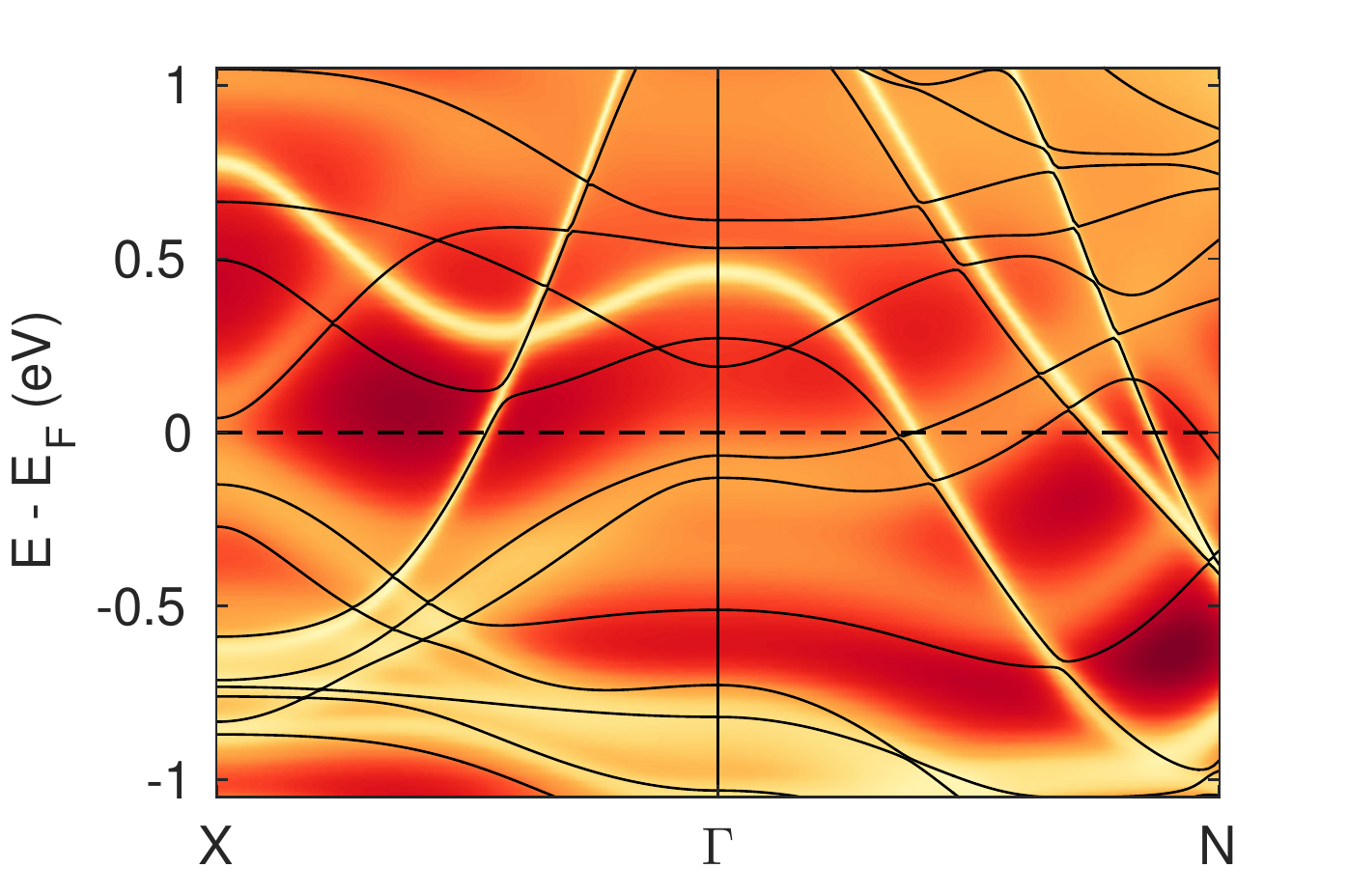}
    }
\caption{Spectral functions from SPRKKR in the CPA and electronic bands from FPLO in the VCA for (Fe$_{1-x}$Co$_{x}$)$_2$B. In (a) the main spin and $d$-orbital character of the two lowest unoccupied and two highest occupied states at $\Gamma$ are indicated with $\downarrow$ denoting the minority spin channel.}
\label{spectralfunctions}
\end{figure*}

For further insight into the MAE of the alloy system we have plotted the Bloch spectral functions from SPRKKR-CPA calculations together with the band structure from FPLO-VCA calculations for a few alloy concentrations in Fig.~\ref{spectralfunctions}. In addition to the smearing in the spectral functions, which can be seen to increase somewhat with $x$, certain bands appear slightly shifted in comparing the bands and spectral functions. These shifts are expected to be mainly due to FP effects, which are absent in SPRKR-CPA, a conclusion which is supported by comparison of Fig.~\ref{spectralfunctions} to Fig.~\ref{Fe2B_bands}, and in general the electronic structure appears very similar in the two models. This similarity is the reason why the VCA and CPA tend to yield qualitatively similar results while the quantitative differences appear because the band smearing is absent in the VCA.
In Ref.~\cite{belashchenko} the filling of states with alloy concentration was investigated and related to the variation of the MAE. It was suggested\cite{belashchenko} that the spin-diagonal part of the spin-orbit operator is most important with rather strong and nearly constant, positive MAE contributions coming from the coupling between majority spin states while coupling between minority spin states yields the variation in the MAE. More specifically, it was suggested that minority spin bands yield a negative contribution for $x \leq 0.3$, which is suppressed as the two minority spin bands on opposite sides of E$_\text{F}$ at the $\Gamma$ point for $x=0.2$ become occupied, which they both are at $x=0.3$. Identification of the character of the bands in Fe$_2$B using a scalar relativistic, spin polarized calculation neglecting SOC supports that the two relevant lowest unoccupied bands at $\Gamma$ in Fig.~\ref{x10bsf} are mainly of minority spin character with magnetic quantum number $m=\pm 1$. It is also confirmed that the two highest occupied states at $\Gamma$ are of minority spin character with magnetic quantum number $m=\pm 2$. Since the spin-diagonal part of the spin-orbit coupling between states of different magnetic quantum numbers yiels a negative MAE contribution, this coincides with the picture given in Ref.~\cite{belashchenko}. However, for a more complete picture of the total effect on the MAE, contributions from bands around the Fermi energy in the entire Brillouin zone should be carefully considered. This is illustrated further by the observation that the qualitative description, just discussed to explain why there is a maximum in the MAE for $x=0.3$, appears to hold also for the FPLO VCA calculations even though these do not show the maximum MAE until $x=0.4$. We finalize this discussion by concluding that the MAE is a delicate property sensitive to the electronic structure in a region near the Fermi energy while major changes tend to be traceable to bands crossing the Fermi energy.

\subsection{Electron Correlations in Fe$_2$B and Co$_2$B within DMFT} 

The disagreement between the presented DFT calculations and experimental measurements regarding the magnetic moment of Co$_2$B motivates further investigation to resolve this issue and correlation effects beyond standard DFT methods will therefore be considered. In order to investigate correlation effects on the magnetic properties of the Fe$_2$B and Co$_2$B compounds, we have performed electronic structure calculations within dynamical mean-field theory (DMFT),\cite{Georges:1996hva, Kotliar:2006fl} using the so-called DFT+DMFT approach\cite{Kotliar:2004ea, DiMarco:2009uu, Lichtenstein1988, Granas2012} as implemented in the full-potential linear muffin-tin orbital (FP-LMTO) code ``RSPt''.\cite{Andersen:1975kh, Wills:2010ej}
The \textit{3d} subset of correlated orbitals was defined using the ``muffin-tin head'' (MT) projection. For further details we refer the reader to Refs. \onlinecite{Grechnev:2007en, DiMarco:2009ea, DiMarco:2009uu}. The DMFT calculations are performed at a finite temperature set to room temperature in this study. Spin-orbit coupling in the muffin-tin region was included in the calculations.
The impurity problem occurring within DMFT was solved via the spin-polarized \textit{T}-matrix fluctuation-exchange (SPTF) solver,\cite{Pourovskii:2005km, DiMarco:2012wr} which is based on perturbation theory.
The Hubbard \textit{U} was set to 1.5 eV and 2.5 eV for Fe$_2$B and Co$_2$B, respectively, and the Hund's exchange \textit{J} to 0.9 eV. These values of $U$ are chosen since values in the range $1-3~\text{eV}$ are commonly used for metallic Fe and Co while using a somewhat larger value for Co than for Fe has been shown to yield a good description of the orbital magnetism in these elements\cite{Chadov2008}. 
The around mean-field approximation was used for the double counting correction.

\begin{figure}[t]
a)\includegraphics[width=0.24\textwidth]{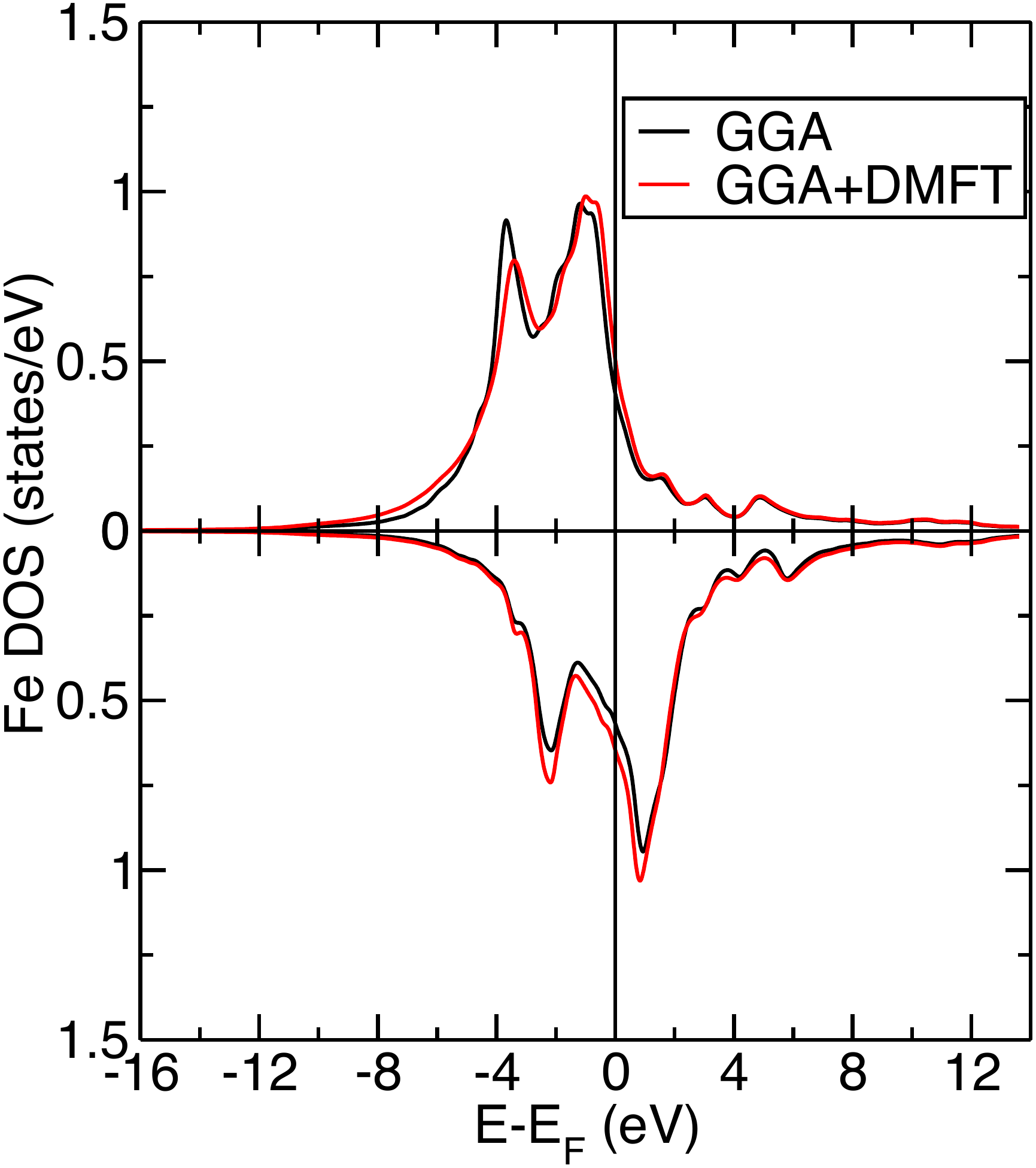}%
b)\includegraphics[width=0.24\textwidth]{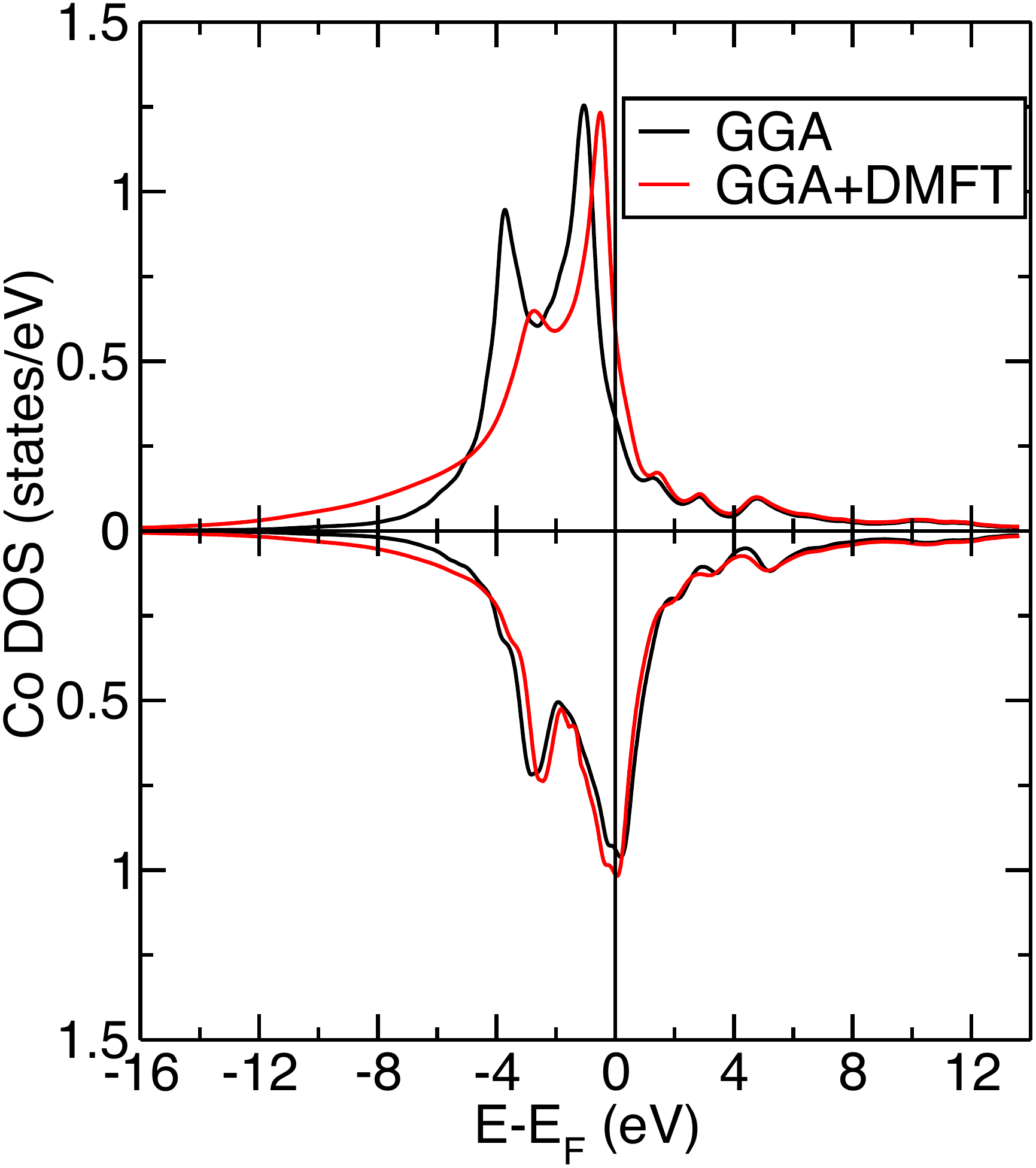}%
\caption{\label{fig:dmft}Projected density of states of the \textit{3d} states in Fe$_2$B (fig. a) and Co$_2$B (fig. b), respectively, calculated within DFT and DMFT (SPTF solver). The Fermi level is set to zero.}
\end{figure} 

\begin{table}[b]
\caption{\label{tab:dmft}
Calculated magnetic moments within DFT and DMFT (SPTF solver).}
\begin{ruledtabular}
\begin{tabular}{lcc}
Co$_2$B & Co spin moment [$\rm \mu_B$] & Co orbital moment [$\rm \mu_B$]\\
DFT & 1.083 & 0.036\\
DMFT & 0.813 & 0.034\\
\hline
Fe$_2$B & Fe spin moment [$\rm \mu_B$] & Fe orbital moment [$\rm \mu_B$]\\
DFT &1.938 & 0.041\\
DMFT & 1.841 & 0.041\\
\end{tabular}
\end{ruledtabular}
\end{table} 
The calculated magnetic moments within DFT and DMFT are given in Table \ref{tab:dmft}. DMFT brings a decrease in the spin polarization for the transition metal moments in both Co$_2$B and Fe$_2$B. The decrease in the Co spin polarization is considerable, in good agreement with the experimentally measured moments. The decrease can be understood by investigating the projected density of states of the transition metals, shown in Fig. \ref{fig:dmft}. Negligible differences among DFT and DMFT are found for the minority spins for both compounds, as well as the majority spins for Fe$_2$B. As for the Co$_2$B majority spins, the peak below the Fermi level is pushed to higher energies, thus decreasing the overall spin polarization. In addition, a satellite-like feature is observed at low energies.

Based on these results we conclude that not only an accurate description of the electrostatic potential, but also a treatment of correlation effects more advanced than that offered by the GGA is of importance to obtain a satisfying picture of the magnetic properties of Co$_2$B. For Fe$_2$B on the other hand, such effects appear to be of less significance. 

\section{Substitutional Doping by 5$d$ Transition Elements}\label{sec:imp}

Various ways of enhancing magnetocrystalline anisotropy in 3$d$ magnets have been discussed by Kuz'min \textit{et al.} \cite{Kuzmin2014}. In this work we tried to prove the concept of $3d$--$5d$ interactions in a more complex system like (Fe$_{0.7}$Co$_{0.3}$)$_2$B. As spin-orbit coupling is essential for the MAE, heavier elements with large SOC can possibly have significant effects on the MAE. Even non-magnetic elements can be of great importance via hybridization\cite{Andersson2007, Bhandary}. Hence, the possibility of tailoring the MAE by adding small amounts of 5$d$ elements substituting some of the Fe and Co in the concentration around (Fe$_{0.7}$Co$_{0.3}$)$_2$B, where the MAE is large and positive, has been explored. 

\subsection{Theory}

\begin{figure}[hbt]
	\centering
	\includegraphics[width=8.5cm]{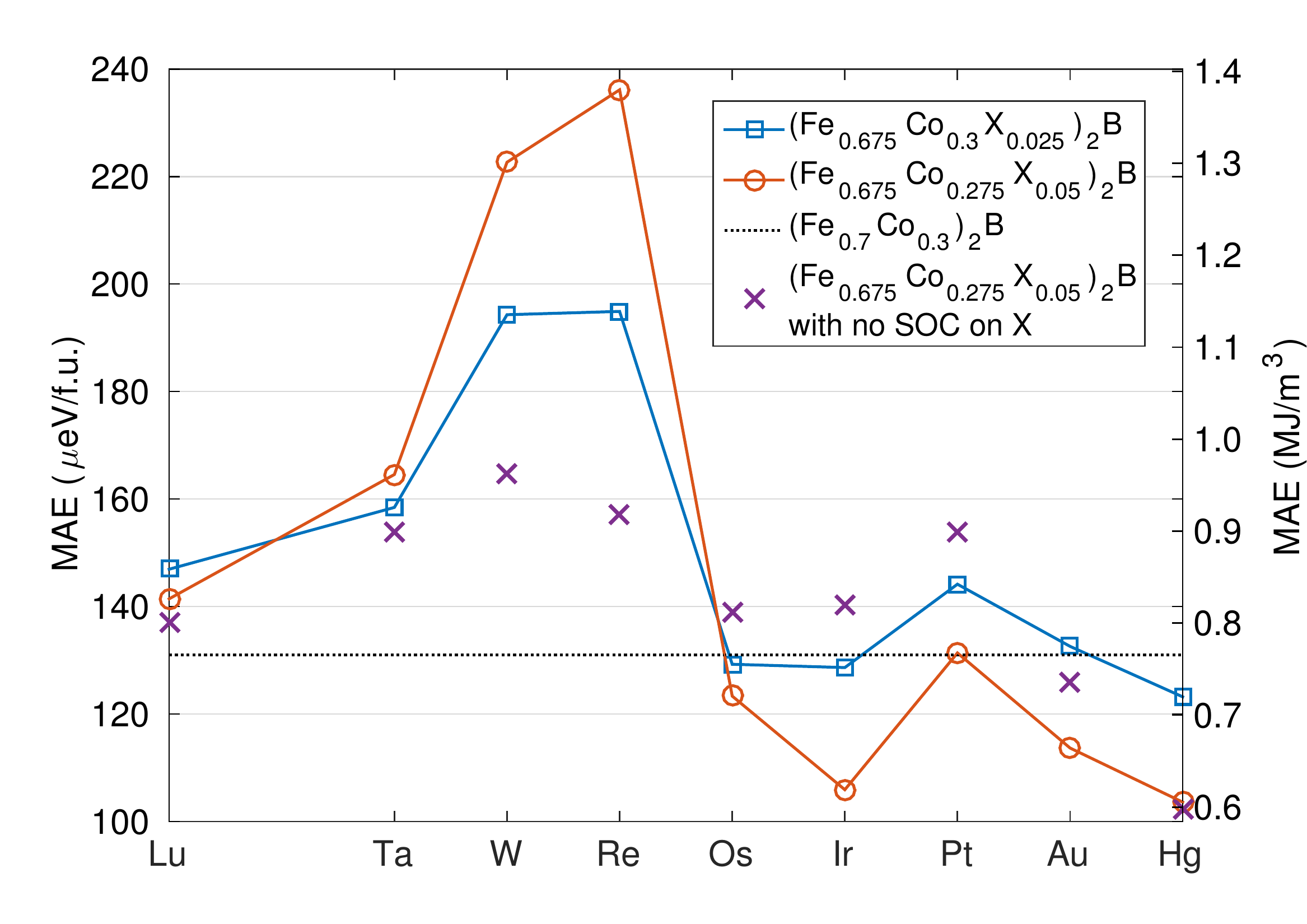}
	\caption{MAE for various elements $X$ in (Fe$_{0.675}$Co$_{0.3}X_{0.025}$)$_2$B and (Fe$_{0.675}$Co$_{0.275}X_{0.05}$)$_2$B. The dotted line indicates the MAE of (Fe$_{0.7}$Co$_{0.3}$)$_2$B for comparison. Crosses indicate the MAE obtained when the spin-orbit coupling on the 5$d$ elements is set to zero.}
	\label{CPAofZ}
\end{figure}

Using the same computational methods as described in Sec.~\ref{CPAsec} (SPRKKR-ASA in the CPA), the magnetic properties and in particular the MAE, was calculated for various elements $X$ from the $5d$ row of the periodic table in the compounds (Fe$_{0.675}$Co$_{0.3}X_{0.025}$)$_2$B and (Fe$_{0.675}$Co$_{0.275}X_{0.05}$)$_2$B. The result is shown in Fig.~\ref{CPAofZ} with the $X$ marked on the $x$-axis and a dotted line indicating the MAE of (Fe$_{0.7}$Co$_{0.3}$)$_2$B for comparison. The calculations indicate that doping with W or Re indeed appears to cause a remarkable increase of the MAE, with the cost of a small decrease in $M_\text{s}$ due to substituting some magnetic elements for non-magnetic ones. The greatest increase in MAE is for Re where there is a rise from $\text{MAE} = 0.77~\text{MJ/m}^3$ to $\text{MAE} = 1.14~\text{MJ/m}^3$ or $1.38~\text{MJ/m}^3$, assuming 2.5~\% or 5.0~\% substitutions per Fe/Co atom, respectively.

The large increase in MAE only appears to be seen when doping with W or Re and not with other elements studied, in agreement with qualitative prediction based on Fig.~\ref{feco2b_color_map}. The MAE is sensitive to the electronic structure around the Fermi energy and the dopants will not only contribute with stronger SOC but also affect the general electronic structure. In some cases this effect will be beneficial for the MAE, such as for W or Re here, and in some cases not, such as for Ir or Ag where the MAE is reduced. In order to investigate the importance of the strong SOC of the 5$d$ atoms, calculations were performed where the SOC is set to zero on these atoms and the results are included in Fig.~\ref{CPAofZ}. There is a small variation in the MAE also in this case where the SOC is zero on the 5$d$ atoms and the trend in this variation is essentially the same as in the case with all SOC included. However, it is clear that these variations can be significantly enhanced due to the strong 5$d$ SOC and that this is what yields the immense MAE for Re or W doping. Other relevant changes to the electronic structure around the Fermi energy caused by the dopants could in principle be analyzed in the spectral functions but it is difficult to observe specific changes due to the small amounts of dopants in the rather complicated spectral functions for the disordered systems and hence these have not been included. As the effect of the dopants is more complex than simply causing an enhancement of the MAE, it could be of interest to study $5d$ substitution around other values of $x$ than $x=0.3$ as well.

\subsection{Experiments on (Fe$_{0.675}$Co$_{0.3}$X$_{0.025}$)$_2$B}

\begin{figure}[hbt]
	\centering
	\includegraphics[width=8.5cm]{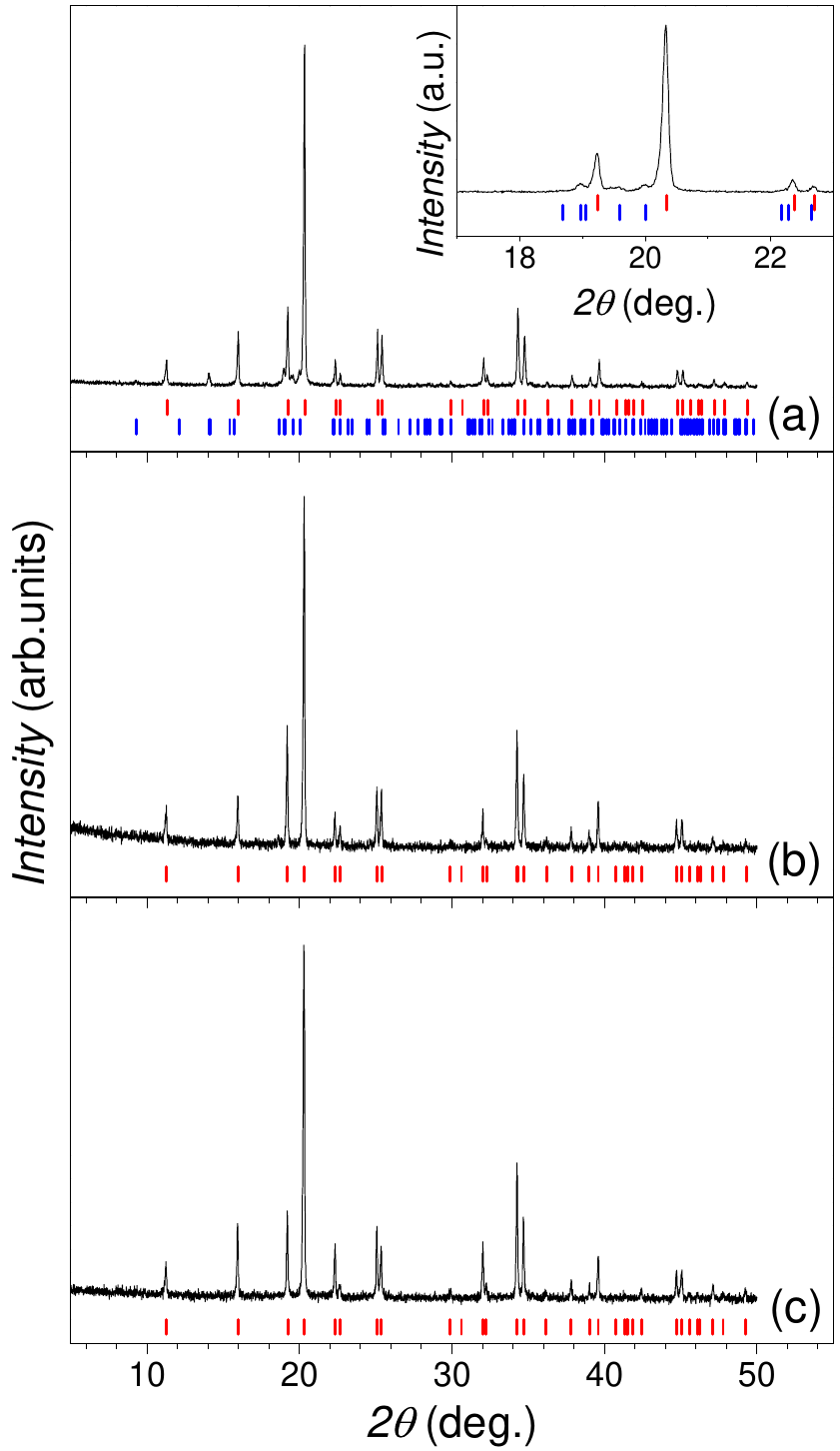}
	\caption{Room temperature XRD plots of (Fe$_{0.675}$Co$_{0.3}X_{0.025}$)$_2$B for $X =$ (a) W, (b) Ir and (c) Re. Colored lines indicate the Bragg positions of Fe$_2$B-phase and CoWB-phase in red and blue, respectively. The inset in (a) shows an enlarged region of $17^\circ-23^\circ$ for W substituted sample.}
	\label{XRD_5d}
\end{figure}

As predicted by the theoretical calculations we tried to substitute the pure (Fe$_{0.7}$Co$_{0.3}$)$_2$B system with 5$d$ elements of W, Re and Ir. The room temperature XRD measurements of 5$d$ doped samples are shown in Fig.~\ref{XRD_5d}.

The room temperature XRD pattern (Fig.~\ref{XRD_5d}a) of W doped sample shows two different phases, main (Fe,Co)$_2$B phase and a secondary CoWB (Pnma) phase. We tried different synthesis methods to avoid the formation of the secondary phase, but in all cases W tends to create a secondary phase rather than replacing the Fe in the main phase. In contrary to the W substitution, single phase materials were obtained for Ir (Fig.~\ref{XRD_5d}b) and Re (Fig.~\ref{XRD_5d}c) substitutions. 

\begin{figure}[hbt]
	\centering
	\includegraphics[width=8.5cm]{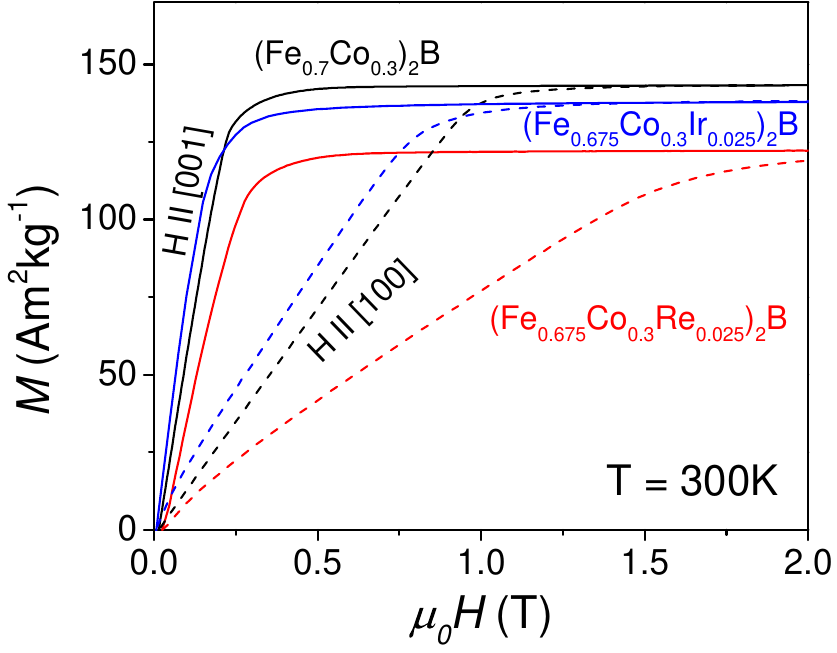}
	\caption{Magnetization curves of (Fe$_{0.7}$Co$_{0.3}$)$_2$B and (Fe$_{0.675}$Co$_{0.3}X_{0.025}$)$_2$B single crystals along [100] and [001] ($X =$ Re, Ir).}
	\label{MH_Re}
\end{figure}

The comparison of the room temperature magnetic measurements of pure (Fe$_{0.7}$Co$_{0.3}$)$_2$B and $5d$ substituted (Fe$_{0.675}$Co$_{0.3}X_{0.025}$)$_2$B with $X =$ Re, Ir samples are shown in Fig.~\ref{MH_Re}. For the Re substituted sample, the saturation magnetization at 2~T decreases from 143.3 Am$^2$kg$^{-1}$ to 122.2 Am$^2$kg$^{-1}$ which is expected due to reduced amount of magnetic elements and, consequently, increased Fe-Fe interaction distances. Contrary to the observed decrease in the saturation magnetization, quite a big improvement is observed in the anisotropy field. The Re substitution increased the anisotropy field of the pure system from 1~T to 1.6~T.

The Ir substitution slightly decreases the saturation magnetization at 2~T to 138.2~Am$^2$kg$^{-1}$ (Fig.~\ref{MH_Re}). In addition the anisotropy field is also decreased down to 0.8~T, which leads to a decrease in the leading anisotropy constant.

\begin{figure}[hbt]
	\centering
	\includegraphics[width=8.5cm]{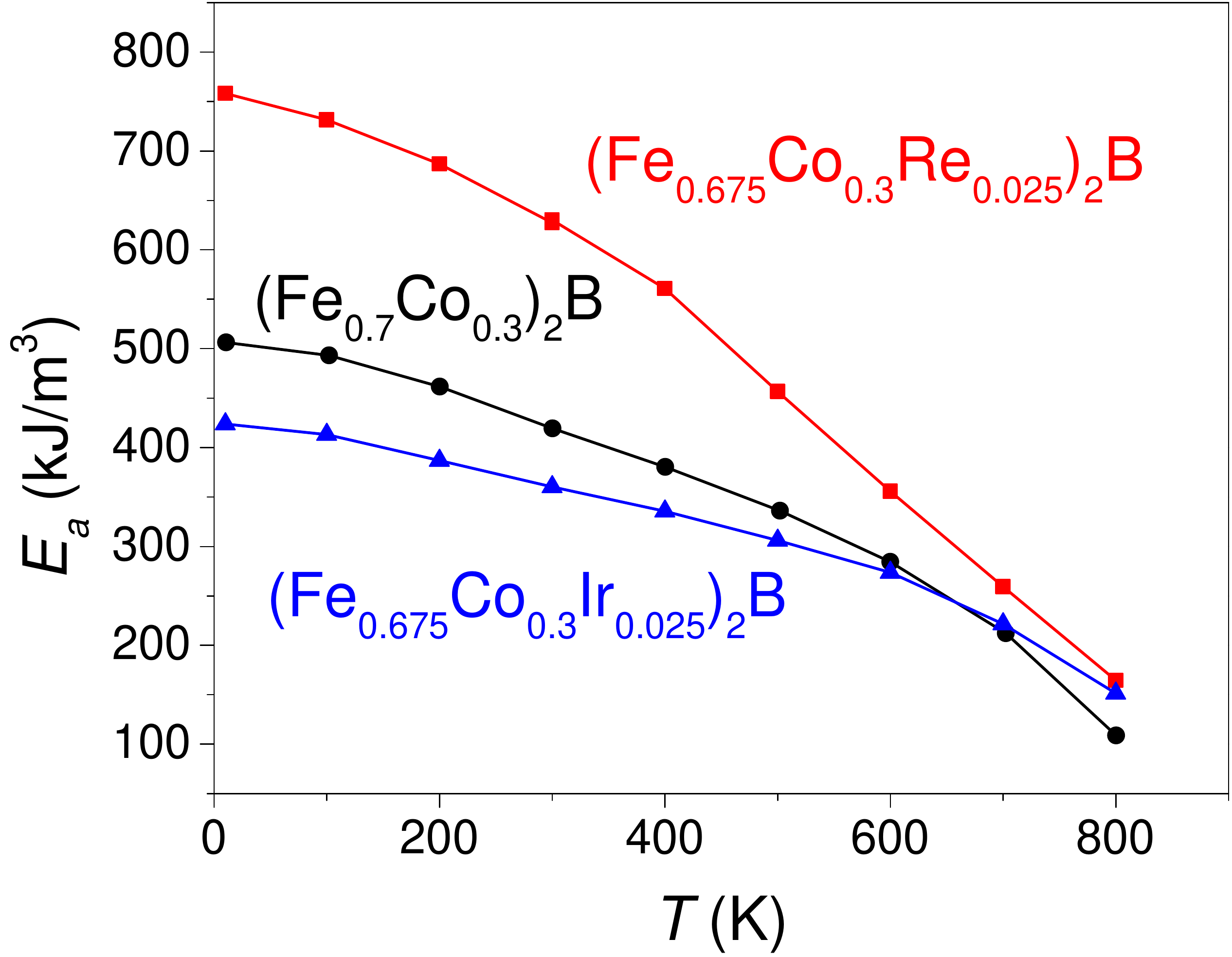}
	\caption{Temperature dependence of the leading anisotropy constants of (Fe$_{0.7}$Co$_{0.3}$)$_{2}$B, (Fe$_{0.675}$Co$_{0.3}$Re$_{0.025}$)$_2$B and (Fe$_{0.675}$Co$_{0.3}$Ir$_{0.025}$)$_2$B.}
	\label{Ea_Re}
\end{figure}

MAE values for both pure and $5d$ substituted samples are shown in Fig.~\ref{Ea_Re} for different temperatures. The results show that for the measured temperature interval an increase in MAE is observed for the Re substitution. The room temperature MAE increases from 0.42~MJm$^{-3}$ to 0.63~MJm$^{-3}$ for Re substitution (approx. 32\% increase). This ratio stays almost constant down to 10~K. This relative increase is similar to the theoretical data shown in Fig.~\ref{CPAofZ}.

The MAE for the Ir substituted sample is 0.36~MJm$^{-3}$ at room temperature, which is approximately 15\% lower than for the pure system. For the high temperature region ($T>600$~K), On the other hand, the MAE shows higher values compared to the pure system.

\section{Conclusions}

In summary, we have presented a combined theoretical and experimental study of structural and magnetic properties of \fecob{} alloys with emphasis on the magnetocrystalline anisotropy as this property is essential, for example, for replacement material candidates for permanent magnet applications. The qualitative shape of the MAE curve as function of Co-concentration agrees with experiment for full-potential calculations in the VCA. However, the quantitative discrepancy is severe as commonly observed when studying the MAE in the VCA\cite{Turek2012, Neise2011, fecoc}. Calculations in the atomic sphere approximation, treating disorder in the CPA, solves the problem of the quantitative disagreement with experiment on the Fe-rich side of the alloy but instead yields a qualitatively incorrect behavior on the Co-rich side. Furthermore, DFT with the generalized gradient approximation for exchange-correlation potential results in an overestimate, compared to experiment, of the magnetic moment of Co$_2$B, which is shown to strongly affect the MAE. It is shown that a reduction of the exchange splitting and the correct magnetic moment of Co$_2$B is reproduced if electron correlations are treated by dynamical mean field theory. This leads us to believe that it would in principle be possible to correctly describe both the magnetic moment and the MAE over the whole range of alloy concentrations by taking into account both full potential effects  and dynamical mean field theory while describing the alloy with the coherent potential approximation. Unfortunately, such calculations are outside of our current capabilities. This provides an alternative description of the observed MAE curve compared to a recent work\cite{belashchenko}, where authors artificially reduce the exchange interaction in order to describe the upturn in MAE on the Co-rich side.
In addition, we have studied the effect of doping these materials with $5d$ elements on their magnetic properties. We have found this to be a viable route to enhancing the MAE of \fecob{} alloys. Simulations suggest that 5~at.\% doping by W or Re should approximately double their MAE, with only modest reduction of saturation magnetization. Experiments support this finding for Re, while synthesis of W-doped \fecob{} was not successful due to formation of additional phases.

\acknowledgments

We acknowledge support from EU-project REFREEPERMAG, the Swedish Research Council, the KAW foundation, ERC grant 247062 (ASD), STANDUPP and eSSENCE. Swedish National Infrastructure for Computing (SNIC) is acknowledged for computer resources. O. Gutfleisch gratefully acknowledges the financial support by the German Federal State of Hessen through its excellence programme LOEWE "RESPONSE".  D.~I. would like to thank I. Di Marco for valuable discussions.


\begin{thebibliography}{99}
\bibitem{kirchmayr} H. R. Kirchmayr, J. Phys. D: Appl. Phys. {\bf 29}, 2763 (1996).
\bibitem{sugimoto} S. Sugimoto, J. Phys. D: Appl. Phys. {\bf 44}, 064001 (2011).
\bibitem{mccallum} R. W. McCallum, L. H. Lewis, R. Skomski, M. J. Kramer, and I. E. Anderson, Annual Review of Materials Research {\bf 44}, 451 (2014).
\bibitem{leitao} J. V. Leitao, M. van der Haar, A. Lefering, E. Br\"{u}ck, J. Mag. Magn. Mat. \textbf{344}, 49 (2013).
\bibitem{ernafe2p} E. K. Delczeg-Czirjak, L. Delczeg, M. P. J. Punkkinen, B. Johansson, O. Eriksson, and L. Vitos, Phys. Rev. B \textbf{82}, 085103 (2010).
\bibitem{Fujii} H. Fujii, T. Hokabe, T. Kamigaichi and T. Okamoto, J. Phys. Sosc. Jpn. {\bf 43}, 41 (1977).
\bibitem{FeNi} Y. Miura, S. Ozaki, Y. Kuwahara, M. Tsujikawa, K. Abe and M. Shirai, J. Phys. Cond. Matt. {\bf 25}, 106005 (2013).
\bibitem{fe16n2} T. K. Kim, M. Takahashi, Appl. Phys. Lett., \textbf{20}, 492 (1972).
\bibitem{heuslers} T. Graf, C. Felser, S. S. P. Parkin, Prog. Solid State Chem. \textbf{39}, 1 (2011).
\bibitem{Burkert2004} T. Burkert, L. Nordstr\"om, O. Eriksson, and O. Heinonen, Phys. Rev. Lett. {\bf 93}, 027203 (2004).
\bibitem{Costa} M. Costa, O. Gr\aa{}n\"{a}s, A. Bergman, P. Venezuela, P. Nordblad, M. Klintenberg, and O. Eriksson, Phys. Rev. B {\bf 86}, 085125 (2012).
\bibitem{Andersson2006} G. Andersson, T. Burkert, P. Warnicke, M. Bj\"orck, B. Sanyal, C. Chacon, C. Zlotea, L. Nordstr\"om, P. Nordblad, and O. Eriksson, Phys. Rev. Lett. {\bf 96}, 037205 (2006).
\bibitem{Warnicke2007} P. Warnicke, G. Andersson, M. Bj\"orck, J. Ferr\'e and P. Nordblad, J. Phys.: Cond. Mat. {\bf 19}, 226218 (2007).
\bibitem{Yildiz2009} F. Yildiz, M. Przybylski, X.-D. Ma, and J. Kirschner, Phys. Rev. B {\bf 80}, 064415 (2009).
\bibitem{Turek2012} I. Turek, J. Kudrnovsk\'{y}, and K. Carva, Physical Review B {\bf 86}, 174430 (2012).
\bibitem{Neise2011} C. Neise, S. Sch\"{o}necker, M. Richter, K. Koepernik, and H. Eschrig, Physica Status Solidi (B) {\bf 248}, 2398 (2011).
\bibitem{fecoc} E. K. Delczeg-Czirjak, A. Edstr\"{o}m, M. Werwi\'{n}ski, J. Rusz, N. V. Skorodumova, L. Vitos, and O. Eriksson, Phys. Rev. B \textbf{89}, 144403 (2014).
\bibitem{fecocexp} L. Reichel, G. Giannopoulos, S. Kauffmann-Weiss, M. Hoffmann, D. Pohl, A. Edstr\"{o}m, S. Oswald, J. Rusz, L. Schultz, and S. Faehler, J. Appl. Phys. \textbf{116}, 213901 (2014).
\bibitem{Bhandary} S. Bhandary, O. Gr\aa{}n\"{a}s, L. Szunyogh, B. Sanyal, L. Nordstr\"{o}m, O. Eriksson, Phys. Rev. B {\bf 84}, 092401 (2011).
\bibitem{l10} A. Edstr\"{o}m, J. Chico, A. Jakobsson, A. Bergman, and J. Rusz, Phys. Rev. B \textbf{90}, 014402 (2014).
\bibitem{mnbi1} V. P. Antropov, V. N. Antonov, L. V. Bekenov, A. Kutepov, G. Kotliar, Phys. Rev. B {\bf90}, 054404 (2014).
\bibitem{mnbi2} J. Cui, J. P. Choi, G. Li, E. Polikarpov, J. Darsell, M. J. Kramer, N. A. Zarkevich, \emph{et al.}, J. Appl. Phys., {\bf115}(17), 17A743 (2014). 
\bibitem{mnbi3} J. Cui, J. P. Choi, G. Li, E. Polikarpov, J. Darsell, N. Overman, M. Olszta, \emph{et al.}, J. Phys.: Cond. Matt., {\bf26}(6), 064212 (2014).
\bibitem{mnbi4} N. V. Rama Rao, M. Gabay, and G. C. Hadjipanayis, J. Phys. D: Appl. Phys., {\bf46}(6), 062001 (2013).
\bibitem{iga} A. Iga, Japan. J. Appl. Phys. \textbf{9}, 415 (1970).
\bibitem{Coene} W.~Coene, F.~Hakkens, R.~Coehoorn, D.~B.~De Mooij, C.~De Waard, J.~Fidler, and R.~Gr\"{o}ssinger, Journal of Magnetism and Magnetic Materials, 96, 189-196 (1991).
\bibitem{Kuzmin2014} M.~D.~Kuzmin, K.~P.~Skokov, H.~Jian, I.~Radulov and O.~Gutfleisch, J. Phys.: Cond. Matt. {\bf26}(6), 064205 (2014). 
\bibitem{Andersson2007} C. Andersson, B. Sanyal, O. Eriksson, L. Nordstr\"{o}m, O. Karis, D. Arvanitis, T. Konishi, E. Holub-Krappe, and J. H. Dunn, Phys. Rev. Lett. {\bf 99}, 177207 (2007).
\bibitem{Belov1956} K.~P.~Belov and A.~N.~Goryaga, Fiz. Met. Metalloved., {\bf2}(1), 3-9 (1956).
\bibitem{Arrott1957} A. Arrott, Physical Review, {\bf108}(6), 1394-1396 (1957).
\bibitem{Blaha01} P. Blaha, K. Schwarz, G. Madsen, D. Kvasnicka and J. Luitz, WIEN2k, An Augmented Plane Wave + Local Orbitals Program for Calculating Crystal Properties (Karlheinz Schwarz, Techn. Universit\"{a}t Wien, Austria), 2001. ISBN 3-9501031-1-2.
\bibitem{Perdew1996} J.P. Perdew, K. Burke, and M. Ernzerhof, Phys. Rev. Lett. \textbf{77}, 3865 (1996).
\bibitem{Cadeville75} M.C. Cadeville and I. Vincze, J.~Phys.~F:~Met.~Phys. \textbf{5}, 790 (1975).
\bibitem{Koepernik99} K.~Koepernik and H.~Eschrig, Phys.~Rev.~B \textbf{59}, 1743 (1999).
\bibitem{Bruno1989} P. Bruno, Phys. Rev. B 39, 865(R) (1989).
\bibitem{lsda} J. P. Perdew, Y. Wang, Phys. Rev. B \textbf{46}, 12947 (1992).
\bibitem{kapfenberger} C. Kapfenberger, B. Albert, R. P\"{o}ttgen, H. Huppertz, Z. F\"{u}r Krist. \textbf{221}, (5-7/2006) (2006).
\bibitem{takacs} L. Takacs, M. C. Cadeville, I. Vincze, J. Phys. F Met. Phys. \textbf{5}, 800 (1975).
\bibitem{sprkkr} H. Ebert, D. K\"{o}dderitzsch, and J. Min\'{a}r, Rep. Prog. Phys. {\bf 74}, 096501 (2011).
\bibitem{Ebert2012} H.~Ebert, ``{The Munich SPR-KKR package, version 6.3},'' 2012.
\bibitem{Wang1996} X.~Wang, R.~Wu, D.~S.~Wang, and A.~J.~Freeman, Phys. Rev. B, \textbf{54}, 61 (1996).
\bibitem{belashchenko} K.~D.~Belashchenko, L.~Ke, M.~D\"{a}ne, L.~X.~Benedict, T.~N.~Lamichhane, V.~Taufour, A.~Jesche, S.~L.~Bu\'{d}ko, P.~C.~Canfield, V.~P.~Antropov, Applied Physics Letters \textbf{106}, 062408 (2015).
\bibitem{Kondorskii} E.~I.~Kondorskii, and E.~Straube, Journal of Experimental and Theoretical Physics, \textbf{63}, 188-193 (1973).
\bibitem{Wu} R. Wu, and A. J. Freeman, Journal of Magnetism and Magnetic Materials, \textbf{200}, 498-514 (1999).
\bibitem{Abate1965} E.~Abate and M.~Asdente, Physical Review \textbf{140}, A1303 (1965).
\bibitem{Liechtenstein} A.~I.~Liechtenstein, M.~I.~Katsnelson, V.~P.~Antropov, and V.~A.~Gubanov, Journal of Magnetism and Magnetic Materials {\bf 67}, 65-74 (1987).
\bibitem{Mohn} Peter Mohn, Magnetism in the Solid State - An Introduction , Springer-Verlag (2003).
\bibitem{CallenCallen} H.~B.~Callen and E.~Callen, The present status of the temperature dependence of magnetocrystalline anisotropy, and the $l(l+1)/2$ power law. Journal of Physics and Chemistry of Solids \textbf{27}, 1271-1285 (1966).
\bibitem{Chadov2008} S. Chadov, H. Minar, M. I. Katsnelson, H. Ebert, D. K\"{o}dderitzsch, and A. I. Lichtenstein, Orbital Magnetism in Transition Metal System: The Role of Local Correlation Effects. EPL \textbf{82}, 37001 (2008).
\bibitem{Jian2015} H. Jian, K. P. Skokov, and O. Gutfleisch, Microstructure and magnetic properties of Mn-Al-C alloy powders prepared by ball milling. Journal of Alloys and Compounds, \textbf{622}, 524-528 (2015)
\bibitem{Ener2015} S. Ener, K. P. Skokov, D. Y. Karpenkov, M. D. Kuzmin, and O. Gutfleisch,  Magnet properties of Mn$_70$Ga$_30$ prepared by cold rolling and magnetic field annealing. Journal of Magnetism and Magnetic Materials, \textbf{382}, 265-270 (2015). 
\bibitem{Gutfleisch2011} O. Gutfleisch, M. A. Willard, E. Br\"uck, C. H. Chen, S. G. Sankar, and J. P. Liu,  Magnetic materials and devices for the 21st century: stronger, lighter, and more energy efficient. Advanced Materials , \textbf{23}, 821-842 (2011).
\bibitem{Lichtenstein1988} A. I. Lichtenstein and M. I. Katsnelson, Ab initio calculations of quasiparticle band structure in correlated systems: LDA++ approach, Physical Review B \textbf{57}, 6884 (1998).
\bibitem{Granas2012} Oscar Gr{\aa}n\"as, Igor Di Marco, Patrik Thunstr\"{o}m, Lars Nordstr\"{o}m, Olle Eriksson, Torbj\"{o}rn Bj\"{o}rkman and JM Wills, Charge self-consistent dynamical mean-field theory based on the full-potential linear muffin-tin orbital method: Methodology and applications, Computational Materials Science {\bf 55}, 295 (2012)
\makeatletter
\providecommand \@ifxundefined [1]{%
 \@ifx{#1\undefined}
}%
\providecommand \@ifnum [1]{%
 \ifnum #1\expandafter \@firstoftwo
 \else \expandafter \@secondoftwo
 \fi
}%
\providecommand \@ifx [1]{%
 \ifx #1\expandafter \@firstoftwo
 \else \expandafter \@secondoftwo
 \fi
}%
\providecommand \natexlab [1]{#1}%
\providecommand \enquote  [1]{``#1''}%
\providecommand \bibnamefont  [1]{#1}%
\providecommand \bibfnamefont [1]{#1}%
\providecommand \citenamefont [1]{#1}%
\providecommand \href@noop [0]{\@secondoftwo}%
\providecommand \href [0]{\begingroup \@sanitize@url \@href}%
\providecommand \@href[1]{\@@startlink{#1}\@@href}%
\providecommand \@@href[1]{\endgroup#1\@@endlink}%
\providecommand \@sanitize@url [0]{\catcode `\\12\catcode `\$12\catcode
  `\&12\catcode `\#12\catcode `\^12\catcode `\_12\catcode `\%12\relax}%
\providecommand \@@startlink[1]{}%
\providecommand \@@endlink[0]{}%
\providecommand \url  [0]{\begingroup\@sanitize@url \@url }%
\providecommand \@url [1]{\endgroup\@href {#1}{\urlprefix }}%
\providecommand \urlprefix  [0]{URL }%
\providecommand \Eprint [0]{\href }%
\providecommand \doibase [0]{http://dx.doi.org/}%
\providecommand \selectlanguage [0]{\@gobble}%
\providecommand \bibinfo  [0]{\@secondoftwo}%
\providecommand \bibfield  [0]{\@secondoftwo}%
\providecommand \translation [1]{[#1]}%
\providecommand \BibitemOpen [0]{}%
\providecommand \bibitemStop [0]{}%
\providecommand \bibitemNoStop [0]{.\EOS\space}%
\providecommand \EOS [0]{\spacefactor3000\relax}%
\providecommand \BibitemShut  [1]{\csname bibitem#1\endcsname}%
\let\auto@bib@innerbib\@empty
\bibitem [{\citenamefont {Georges}\ \emph {et~al.}(1996)\citenamefont
  {Georges}, \citenamefont {Kotliar}, \citenamefont {Krauth},\ and\
  \citenamefont {Rozenberg}}]{Georges:1996hva}%
  \BibitemOpen
  \bibfield  {author} {\bibinfo {author} {\bibfnamefont {A.}~\bibnamefont
  {Georges}}, \bibinfo {author} {\bibfnamefont {G.}~\bibnamefont {Kotliar}},
  \bibinfo {author} {\bibfnamefont {W.}~\bibnamefont {Krauth}}, \ and\ \bibinfo
  {author} {\bibfnamefont {M.~J.}\ \bibnamefont {Rozenberg}},\ }\href {\doibase
  10.1103/RevModPhys.68.13} {\bibfield  {journal} {\bibinfo  {journal} {Rev.
  Mod. Phys.}\ }\textbf {\bibinfo {volume} {68}},\ \bibinfo {pages} {13}
  (\bibinfo {year} {1996})}\BibitemShut {NoStop}%
\bibitem [{\citenamefont {Kotliar}\ \emph {et~al.}(2006)\citenamefont
  {Kotliar}, \citenamefont {Savrasov}, \citenamefont {Haule}, \citenamefont
  {Oudovenko}, \citenamefont {Parcollet},\ and\ \citenamefont
  {Marianetti}}]{Kotliar:2006fl}%
  \BibitemOpen
  \bibfield  {author} {\bibinfo {author} {\bibfnamefont {G.}~\bibnamefont
  {Kotliar}}, \bibinfo {author} {\bibfnamefont {S.~Y.}\ \bibnamefont
  {Savrasov}}, \bibinfo {author} {\bibfnamefont {K.}~\bibnamefont {Haule}},
  \bibinfo {author} {\bibfnamefont {V.~S.}\ \bibnamefont {Oudovenko}}, \bibinfo
  {author} {\bibfnamefont {O.}~\bibnamefont {Parcollet}}, \ and\ \bibinfo
  {author} {\bibfnamefont {C.~A.}\ \bibnamefont {Marianetti}},\ }\href
  {\doibase 10.1103/RevModPhys.78.865} {\bibfield  {journal} {\bibinfo
  {journal} {Rev. Mod. Phys.}\ }\textbf {\bibinfo {volume} {78}},\ \bibinfo
  {pages} {865} (\bibinfo {year} {2006})}\BibitemShut {NoStop}%
\bibitem [{\citenamefont {Kotliar}\ and\ \citenamefont
  {Vollhardt}(2004)}]{Kotliar:2004ea}%
  \BibitemOpen
  \bibfield  {author} {\bibinfo {author} {\bibfnamefont {G.}~\bibnamefont
  {Kotliar}}\ and\ \bibinfo {author} {\bibfnamefont {D.}~\bibnamefont
  {Vollhardt}},\ }\href {\doibase 10.1063/1.1712502} {\bibfield  {journal}
  {\bibinfo  {journal} {Phys. Today}\ }\textbf {\bibinfo {volume} {57}},\
  \bibinfo {pages} {53} (\bibinfo {year} {2004})}\BibitemShut {NoStop}%
\bibitem [{\citenamefont {Di~Marco}(2009)}]{DiMarco:2009uu}%
  \BibitemOpen
  \bibfield  {author} {\bibinfo {author} {\bibfnamefont {I.}~\bibnamefont
  {Di~Marco}},\ }\emph {\bibinfo {title} {{Correlation effects in the
  electronic structure of transition metals and their compounds}}},\ \href
  {http://repository.ubn.ru.nl/handle/2066/75407} {Ph.D. thesis},\ \bibinfo
  {school} {Radboud University Nijmegen} (\bibinfo {year} {2009})\BibitemShut
  {NoStop}%
\bibitem [{\citenamefont {Andersen}(1975)}]{Andersen:1975kh}%
  \BibitemOpen
  \bibfield  {author} {\bibinfo {author} {\bibfnamefont {O.~K.}\ \bibnamefont
  {Andersen}},\ }\href {\doibase 10.1103/PhysRevB.12.3060} {\bibfield
  {journal} {\bibinfo  {journal} {Phys. Rev. B}\ }\textbf {\bibinfo {volume}
  {12}},\ \bibinfo {pages} {3060} (\bibinfo {year} {1975})}\BibitemShut
  {NoStop}%
\bibitem [{\citenamefont {Wills}\ \emph {et~al.}(2010)\citenamefont {Wills},
  \citenamefont {Eriksson}, \citenamefont {Andersson}, \citenamefont {Delin},
  \citenamefont {Grechnyev},\ and\ \citenamefont {Alouani}}]{Wills:2010ej}%
  \BibitemOpen
  \bibfield  {author} {\bibinfo {author} {\bibfnamefont {J.~M.}\ \bibnamefont
  {Wills}}, \bibinfo {author} {\bibfnamefont {O.}~\bibnamefont {Eriksson}},
  \bibinfo {author} {\bibfnamefont {P.}~\bibnamefont {Andersson}}, \bibinfo
  {author} {\bibfnamefont {A.}~\bibnamefont {Delin}}, \bibinfo {author}
  {\bibfnamefont {O.}~\bibnamefont {Grechnyev}}, \ and\ \bibinfo {author}
  {\bibfnamefont {M.}~\bibnamefont {Alouani}},\ }\href {\doibase
  10.1007/978-3-642-15144-6} {\emph {\bibinfo {title} {{Full-Potential
  Electronic Structure Method}}}},\ \bibinfo {series} {Springer Series in
  Solid-State Sciences}, Vol.\ \bibinfo {volume} {167}\ (\bibinfo  {publisher}
  {Springer Berlin Heidelberg},\ \bibinfo {address} {Berlin, Heidelberg},\
  \bibinfo {year} {2010})\BibitemShut {NoStop}%
\bibitem [{\citenamefont {Grechnev}\ \emph {et~al.}(2007)\citenamefont
  {Grechnev}, \citenamefont {Di~Marco}, \citenamefont {Katsnelson},
  \citenamefont {Lichtenstein}, \citenamefont {Wills},\ and\ \citenamefont
  {Eriksson}}]{Grechnev:2007en}%
  \BibitemOpen
  \bibfield  {author} {\bibinfo {author} {\bibfnamefont {A.}~\bibnamefont
  {Grechnev}}, \bibinfo {author} {\bibfnamefont {I.}~\bibnamefont {Di~Marco}},
  \bibinfo {author} {\bibfnamefont {M.}~\bibnamefont {Katsnelson}}, \bibinfo
  {author} {\bibfnamefont {A.}~\bibnamefont {Lichtenstein}}, \bibinfo {author}
  {\bibfnamefont {J.}~\bibnamefont {Wills}}, \ and\ \bibinfo {author}
  {\bibfnamefont {O.}~\bibnamefont {Eriksson}},\ }\href {\doibase
  10.1103/PhysRevB.76.035107} {\bibfield  {journal} {\bibinfo  {journal} {Phys.
  Rev. B}\ }\textbf {\bibinfo {volume} {76}},\ \bibinfo {pages} {035107}
  (\bibinfo {year} {2007})}\BibitemShut {NoStop}%
\bibitem [{\citenamefont {Di~Marco}\ \emph {et~al.}(2009)\citenamefont
  {Di~Marco}, \citenamefont {Min{\'a}r}, \citenamefont {Chadov}, \citenamefont
  {Katsnelson}, \citenamefont {Ebert},\ and\ \citenamefont
  {Lichtenstein}}]{DiMarco:2009ea}%
  \BibitemOpen
  \bibfield  {author} {\bibinfo {author} {\bibfnamefont {I.}~\bibnamefont
  {Di~Marco}}, \bibinfo {author} {\bibfnamefont {J.}~\bibnamefont {Min{\'a}r}},
  \bibinfo {author} {\bibfnamefont {S.}~\bibnamefont {Chadov}}, \bibinfo
  {author} {\bibfnamefont {M.}~\bibnamefont {Katsnelson}}, \bibinfo {author}
  {\bibfnamefont {H.}~\bibnamefont {Ebert}}, \ and\ \bibinfo {author}
  {\bibfnamefont {A.}~\bibnamefont {Lichtenstein}},\ }\href {\doibase
  10.1103/PhysRevB.79.115111} {\bibfield  {journal} {\bibinfo  {journal} {Phys.
  Rev. B}\ }\textbf {\bibinfo {volume} {79}},\ \bibinfo {pages} {115111}
  (\bibinfo {year} {2009})}\BibitemShut {NoStop}%
\bibitem [{\citenamefont {Pourovskii}\ \emph {et~al.}(2005)\citenamefont
  {Pourovskii}, \citenamefont {Katsnelson},\ and\ \citenamefont
  {Lichtenstein}}]{Pourovskii:2005km}%
  \BibitemOpen
  \bibfield  {author} {\bibinfo {author} {\bibfnamefont {L.~V.}\ \bibnamefont
  {Pourovskii}}, \bibinfo {author} {\bibfnamefont {M.~I.}\ \bibnamefont
  {Katsnelson}}, \ and\ \bibinfo {author} {\bibfnamefont {A.~I.}\ \bibnamefont
  {Lichtenstein}},\ }\href {\doibase 10.1103/PhysRevB.72.115106} {\bibfield
  {journal} {\bibinfo  {journal} {Phys. Rev. B}\ }\textbf {\bibinfo {volume}
  {72}},\ \bibinfo {pages} {115106} (\bibinfo {year} {2005})}\BibitemShut
  {NoStop}%
\bibitem [{\citenamefont {Di~Marco}\ \emph {et~al.}(shed)\citenamefont
  {Di~Marco}, \citenamefont {Thunstr{\"o}m}, \citenamefont {Gr{\aa}n{\"a}s},
  \citenamefont {Pourovskii}, \citenamefont {Katsnelson}, \citenamefont
  {Nordstr{\"o}m},\ and\ \citenamefont {Eriksson}}]{DiMarco:2012wr}%
  \BibitemOpen
  \bibfield  {author} {\bibinfo {author} {\bibfnamefont {I.}~\bibnamefont
  {Di~Marco}}, \bibinfo {author} {\bibfnamefont {P.}~\bibnamefont
  {Thunstr{\"o}m}}, \bibinfo {author} {\bibfnamefont {O.}~\bibnamefont
  {Gr{\aa}n{\"a}s}}, \bibinfo {author} {\bibfnamefont {L.}~\bibnamefont
  {Pourovskii}}, \bibinfo {author} {\bibfnamefont {M.~I.}\ \bibnamefont
  {Katsnelson}}, \bibinfo {author} {\bibfnamefont {L.}~\bibnamefont
  {Nordstr{\"o}m}}, \ and\ \bibinfo {author} {\bibfnamefont {O.}~\bibnamefont
  {Eriksson}},\ }\href@noop {} {\enquote {\bibinfo {title} {{An LDA+DMFT study
  of the orbital magnetism and total energy properties of the late transition
  metals: conserving and non-conserving approximations}},}\ } (\bibinfo {year}
  {unpublished})\BibitemShut {NoStop}%

\end{thebibliography}
\end{document}